\documentclass[fleqn,usenatbib]{mnras}

\usepackage{newtxtext,newtxmath}
\usepackage[T1]{fontenc}

\DeclareRobustCommand{\VAN}[3]{#2}
\let\VANthebibliography\thebibliography
\def\thebibliography{\DeclareRobustCommand{\VAN}[3]{##3}\VANthebibliography}

\usepackage[dvipsnames]{xcolor}

\usepackage{graphicx}	% Including figure files
\usepackage{amsmath}	% Advanced maths commands
\usepackage{gensymb}
\title[Mixing, Clouds, and Hot Jupiters]{The Impact of Mixing Treatments on Cloud Modelling in 3D Simulations of Hot Jupiters}

\author[D. A. Christie et al.]{D. A. Christie,$^{1}$\thanks{E-mail: d.christie@exeter.ac.uk}
N. J. Mayne,$^{1}$
S. Lines,$^{1,2}$
V. Parmentier,$^{3}$
J. Manners,$^{1,4}$
I. Boutle,$^{1,2}$
\newauthor B. Drummond,$^{1,2}$ T. Mikal-Evans,$^{5}$ D. K. Sing,$^{6,7}$
and K. Kohary.$^{1}$ 
\\
$^{1}$Physics and Astronomy, College of Engineering, Mathematics and Physical Sciences, University of Exeter, Exeter, EX4 4QL, UK\\
$^{2}$Met Office, FitzRoy Road, Exeter EX1 3PB, UK\\
$^{3}$Atmospheric Oceanic and Planetary Physics, Department of Physics, University of Oxford, OX1 3PU, UK\\
$^{4}$Global Systems Institute, University of Exeter, Level 8 Laver Building, North Park Road, Exeter EX4 4QE, UK  \\
$^{5}$Kavli Institute for Astrophysics and Space Research, Massachusetts Institute of Technology, Cambridge, MA 02139, USA \\
$^{6}$Department of Earth \& Planetary Sciences, Johns Hopkins University, Baltimore, MD, USA\\
$^{7}$Department of Physics \& Astronomy, Johns Hopkins University, Baltimore, MD, USA
}

\date{Accepted XXX. Received YYY; in original form ZZZ}

\pubyear{2021}

\begin{document}
\label{firstpage}
\pagerange{\pageref{firstpage}--\pageref{lastpage}}
\maketitle

\begin{abstract}
We present results of 3D hydrodynamical simulations of HD209458b including a coupled, radiatively-active cloud model ({\sc EddySed}).  We investigate the role of the mixing by replacing the default convective treatment used in previous works with a more physically relevant mixing treatment ($K_{zz}$) based on global circulation.  We find that uncertainty in the efficiency of sedimentation through the sedimentation factor $f_\mathrm{sed}$ plays a larger role in shaping cloud thickness and its radiative feedback on the local gas temperatures -- e.g. hot spot shift and day-to-night side temperature gradient -- than the switch in mixing treatment.  We demonstrate using our new mixing treatments that simulations with cloud scales which are a fraction of the pressure scale height improve agreement with the observed transmission spectra, the emission spectra, and the Spitzer 4.5 $\mathrm{\mu m}$ phase curve, although our models are still unable to reproduce the optical and UV transmission spectra.  We also find that the inclusion of cloud increases the transit asymmetry in the optical between the east and west limbs, although the difference remains small ($\lesssim 1\%$).
\end{abstract}


%%%%%%%%%%%%%%%%%%%%%%%%%%%%%%%%%%%%%%%%%%%%%%%%%%

%%%%%%%%%%%%%%%%% BODY OF PAPER %%%%%%%%%%%%%%%%%%

\section{Introduction}

Clouds and hazes have been found to be common across the range of currently discovered exoplanets. The presence of clouds, or hazes, (hereafter we use the generic term `clouds' to describe non-gas phase opacity sources in the atmosphere) has been inferred from observations of many targets through the wavelength dependence of transmission spectra \citep{2008A&A...481L..83L,2015MNRAS.447..463N} or through muted spectral signatures of the underlying atmosphere \citep{2013ApJ...774...95D,2016Natur.529...59S,2016ApJ...823..109I}. Additionally, shifts in the peak optical or IR flux \citep[see][]{demory_2011, dang_2018}, as a function of orbital phase, and temporal variability \citep{armstrong_2016} have been interpreted as evidence for clouds. Yet despite their occurrence being well studied, the mechanisms determining their presence and persistence remain poorly quantified \citep[e.g.][]{stevenson_2010,kreidberg_2014,2016Natur.529...59S,kreidberg_2018,bruno_2018}.  There is some indication for hotter atmospheres generally having clearer skies on the day side and along the terminator, inferred from negative correlation between equilibrium temperature and muted spectral features \citep{stevenson_2016,heng_2016,fu_2017}, although other parameters such as planet composition may also play a role \citep{2016Natur.529...59S,bruno_2018}. 

%Importantly, no clear correlation of the transmission spectra properties with planetary parameters has been found, and differences in the transmission for planets with similar gravities and equilibrium temperatures suggest planet composition may play a role \citep{2016Natur.529...59S,bruno_2018}. 

Cloud treatments and parametrizations have been included in 1D forward models \citep[e.g.][]{seager_2000,hubbard_2001,brown_2001,2017A&A...600A..10M,charnay_2018,goyal_2018}, used directly to interpret spectra \citep[e.g.][]{2017A&A...600A..10M} or as part of retrieval frameworks to determine the optical structure of the target atmosphere implied by the observations \citep[e.g.][]{Barstow_2017, Wakeford_2017}. These treatments can range from the most simple approach, that of prescribed `decks' of opacity \citep[e.g.][]{Barstow_2017,Fisher_2018,Pinhas_2019}, to parameterisations attempting to capture the basic physics \citep[e.g.,][hereafter termed the {\sc EddySed} model]{2001ApJ...556..872A} to more sophisticated microphysical approaches \citep[e.g.][]{juncher_2017,Powell_2019}. However, there is a growing realisation that 1D models are insufficient to correctly capture the physical state of, in particular, short period exoplanets such as hot Jupiters. For example, works such as \citet{Pluriel_2020} and \citet{tremblin_2017,mayne_2017,drummond_2018a,drummond_2018c,mendonca_2018,drummond_2020} have demonstrated the need for higher dimensionality models, even in the absence of clouds, with work such as \citet{line_2016,Irwin_2020,Taylor_2020,feng_2020} exploring higher dimensionality retrieval methods. Planetary magnetic fields, partially coupled to the gas by ions in the atmosphere, cannot be effectively modelled in one dimension and introduce both drag and heating capable of impacting the atmospheric winds and circulation patterns \citep{rauscher_2013,rogers_2014,hindle_2019}.  Clouds themselves can act to exacerbate departures from symmetry, hindering a collapse to 1D treatments through, for example, day-night side differences in cloud coverage \citep[e.g.][]{Helling_2019} and differences between the east and west terminators \citep{Powell_2019}.

Cloud models implemented in 3D have a similar range of complexity to that found in 1D, from simulation of gas-phase only atmospheres and post--processing to determine cloud properties \citep[e.g.][]{parmentier_2016,helling_2016} to simple, prescribed opacity cloud decks \citep[e.g.][]{dobbs_dixon_2013,mendonca_2018b}, and on to models coupling 3D dynamics and radiative transfer with either parametrised cloud schemes \citep[e.g.][]{roman_2019,harada_2019,lines_2019,parmentier_2020}, or microphysical treatments \citep[e.g.][]{2016A&A...594A..48L,2018A&A...615A..97L,2018MNRAS.481..194L}. Clouds, however, are complex, and currently there is no single approach able to capture the full range of physical mechanisms required to self--consistently predict their presence and impact on observations. Therefore, this range of approaches should be seen as complementary and, indeed, required to gradually determine the key physical processes, and thereby eventually construct a model of the minimal required complexity.

Efforts to couple a microphysical cloud-formation model to 3D global climate models (GCMs) of hot Jupiters include \citet{2016A&A...594A..48L} and \citet{2018A&A...615A..97L}, both of which use the underlying cloud formation model of \citet{2013RSPTA.37110581H}. These studies are generally limited by the computational expense of the underlying chemistry and the long timescales needed to approach a converged solution. In particular, the simulations of \citet{2018A&A...615A..97L} result in the presence of large quantities of small particles at high altitude, which also require extremely long timescales to settle under gravity. These models have, however, shown that the radiative feedback of the clouds themselves plays a significant role in shaping the thermal structure of the atmosphere and must be included \citep[see also][]{roman_2019}. Such limitations have motivated the use of more simplified cloud models in GCMs. Studies using simplified temperature--dependent cloud schemes have demonstrated potential trends across a population of gas giant exoplanets, revealing a transition from cloud--free atmospheres at higher irradiation temperatures ($\gtrsim$3000\,K) to cloud formation on the nightside and western--limb \citep{roman_2020,parmentier_2020}. For HD~209458b, \citet{lines_2019} coupled {\sc EddySed} to the Met Office's {\sc Unified Model} ({UM}) and performed simulations extending to about 500 Earth days. The results of \citet{lines_2019} again revealed the importance of the radiative feedback due to clouds but also highlighted the importance of the particle size distributions and the vertical extent of clouds. However, simulations such as those of \citet{lines_2019} employ a simplified parametrisation of the dynamical mixing in the atmosphere, which, in the specific case of \citet{lines_2019} is based on a consideration of convection. In this paper, we continue the work of \citet{lines_2019} in building a more complex cloud model with a limited increase in computational overhead, substituting the default mixing treatment for a more physically relevant model based on mixing found via 3D simulations \citep{2013A&A...558A..91P}  as well as further investigating the relative impact of the sedimentation efficiency on atmospheric structure and observational diagnostics.   

This paper is laid out as follows, in Section \ref{sec:model} we describe our model formulation, including a brief description of the 3D model, followed by explanation of the main features of the cloud model and mixing treatment, as well as introducing the suite of simulations we use. In Section \ref{sec:results} we then present the resulting atmospheric structure from our simulations, the implications of the different mixing and cloud parameters, before demonstrating the impact on synthetic observations. Finally, we conclude and discuss potential avenues for future progress in Section \ref{sec:conclusions}.

\section{The Model}
\label{sec:model}

In this section we first briefly outline the GCM we use to model the atmosphere of HD209458b, followed by a discussion of the cloud model specifically, before describing the set of simulations used for this study. 

\subsection{The Met Office's {\sc Unified Model}}
\label{subsec:um}
The underlying GCM used to perform the simulations is that of the Met Office, termed the {\sc Unified Model} (UM). The UM's dynamical core, {\sc ENDGame} solves the full, deep–atmosphere  and non-hydrostatic  Navier-Stokes equations \citep[see][for discussion]{2014QJRMS.140.1505W,2014A&A...561A...1M,mayne_2019}. Adaptations of the dynamical core, and the results of benchmarking and testing for hot Jupiters are detailed in \citet{2014A&A...561A...1M,2017A&A...604A..79M}. Radiative transfer within the model is handled using the open-source {\sc SOCRATES} code based on \citet{1996QJRMS.122..689E}. {\sc SOCRATES} has also been isolated, adapted, benchmarked and tested for hot Jupiters in \citet{2014A&A...564A..59A, amundsen_2017}, before coupled radiative--dynamic, cloud--free simulations were performed in \citet{2016A&A...595A..36A}. Gas-phase opacity sources are $\mathrm{H_2O}$, $\mathrm{CO}$,  $\mathrm{CH_4}$,  $\mathrm{NH_3}$,  $\mathrm{Li}$,  $\mathrm{Na}$,  $\mathrm{K}$,  $\mathrm{Rb}$,  $\mathrm{Cs}$  and  $\mathrm{H_2-H_2}$ and  $\mathrm{H_2-He}$ collision induced absorption (CIA) with the opacities computed using the correlated-k method and ExoMol linelists \citep{2012MNRAS.425...21T,2016JMoSp.327...73T}.  The individual gas phase abundances are taken from the analytic fits of \citet{1999ApJ...512..843B} with modifications to Alkali metal abundances outlined in \citet{2016A&A...595A..36A}. The UM has also been coupled to gas--phase chemistry schemes of various sophistication \citep{drummond_2018a,drummond_2018b,drummond_2018c,drummond_2020}, which although not employed in this study have been used to demonstrate the importance of 3D mixing in determining the chemical composition of hot Jupiter atmospheres. Cloud treatments have been added to the UM through both the inclusion of a microphysical model \citep[based on][]{2013RSPTA.37110581H} following the work of \citet{2016A&A...594A..48L}, which was applied by \citet{2018A&A...615A..97L,2018MNRAS.481..194L}, and the parametrised {\sc EddySed} scheme of \citet{2001ApJ...556..872A} applied by \citet{lines_2019}. In \citet{lines_2019} cloud opacities are calculated from pre-computed tables assuming a log-normal distribution in particle sizes. In this work, we essentially build on the study of \citet{lines_2019}, using the same model setup, but altering the treatment of dynamical mixing and slightly expanding the cloud optical treatment. 

\subsection{The {\sc EddySed} Model}
\label{subsec:eddysed} 

%Our intent is the use a one-dimensional cloud model to approximate cloud physics without the need for tracers and chemical networks, while still improving on simplified cloud schemes used elsewhere.  We adopt the {\sc EddySed} model as it allows for sedimentation to be modelled resulting in a parametrised cloud scale height 

{\sc EddySed}, as described in \citet{2001ApJ...556..872A}, is a one-dimensional globally-averaged/horizontally-homogeneous  phase–equilibrium and parametrised cloud model appropriate for substellar atmospheres. The fundamental assumption of the model is that the vertical extent of the cloud is determined by a balance of upward transport of vapour (with mass mixing ratio $q_\mathrm{v}$) and condensate (with mass mixing ratio $q_\mathrm{c}$) via mixing and downward transport of condensate via sedimentation. This is coupled with the assumption that the mass-averaged sedimentation velocity ($\left<v_\mathrm{sed}\right>$) is proportional to a characteristic mixing velocity, $w_\star$, which is equal to the ratio of the vertical mixing parameter (or eddy--diffusion coefficient, $K_\mathrm{zz}$) and mixing length-scale ($L_\mathrm{mix}$) i.e., $w_\star=K_\mathrm{zz}/L_\mathrm{mix}$. This is captured in the governing equation
\begin{equation}
    -K_{zz}\frac{\partial q_\mathrm{t}}{\partial z} = \left<v_\mathrm{sed}\right>q_\mathrm{c} = f_\mathrm{sed} w_\star q_\mathrm{c}\,\,,
    \label{Eqn:Balance}
\end{equation}
where $z$ is the vertical coordinate, $q_\mathrm{t} = q_\mathrm{c} + q_\mathrm{v}$ is the total mixing ratio for a given species, and $f_\mathrm{sed}$ is the sedimentation factor. Equation \ref{Eqn:Balance} is solved separately for each condensable species, and as such clouds of one species do not impact those of another directly; however, as all cloud species interact radiatively with the atmosphere, indirect influence is possible. As the sedimentation velocity is proportional to $K_{zz}$ in equation \ref{Eqn:Balance}, the cloud scale height, defined by $L_\mathrm{cloud} = f_\mathrm{sed}^{-1}L_\mathrm{mix}$ is actually independent of $K_{zz}$.
The condensate mixing ratio $q_\mathrm{c}$ is taken to be the excess of a particular species above the local vapour saturation concentration ($q_\mathrm{vs}$),
\begin{equation}
    q_\mathrm{c} = \max\left(0,q_\mathrm{t} - q_\mathrm{vs}\right)\,\,.
    \label{Eqn:Condense}
\end{equation}
 
\noindent The critical vapour saturation mixing ratios $q_\mathrm{vs}$ are taken from \citet{2001ApJ...556..872A,kozasa_1989,visscher_2006,visscher_2010,morley_2012}. Although there is debate about what species are realistically able to condense into clouds in exoplanetary atmospheres (see \citealt{Powell_2019}), we have adopted a maximalist approach and allowed all possible condensable species in {\sc EddySed} to form clouds.  A list of the sources as well as the assumed mass mixing ratios at the base of the atmosphere ($q_\mathrm{below}$) are given in Appendix \ref{app:sat_vap}. Equations \ref{Eqn:Balance} and \ref{Eqn:Condense} are used to determine the distribution of clouds within the atmosphere for a given pressure and temperature profile.   

The particle size distribution is derived from the condition $\left<v_\mathrm{sed}\right> = f_\mathrm{sed} w_\star$.  Given the characteristic radius $r_w$ where $v_\mathrm{sed}\left(r_w\right) = w_\star$ and assuming the sedimentation velocity scales as $v_\mathrm{sed}\left(r\right) = w_\star\left(r/r_w\right)^\alpha$, the peak radius $r_\mathrm{g}$ of the assumed log-normal distribution can be shown to be

\begin{equation}
    r_\mathrm{g} = r_w f_\mathrm{sed}^{1/\alpha}\exp\left(-\frac{\alpha+6}{2}\ln^2\sigma \right),
    \label{Eqn:rg}
\end{equation}

\noindent where $\sigma$ is the geometric width of the log-normal distribution.   We take $\sigma=2$ for consistency with \citet{lines_2019} and note that this choice for $\sigma$ is used in \citet{2017A&A...600A..10M} and \citet{gao_2018}; however,  microphysical models of cloud formation that employ binned size distributions \citep[e.g.,][]{gao_2018,powell_2018,Powell_2019} consistently find distributions that are broad and often double peaked, corresponding to the separate nucleation and condensation modes, which is potentially at odds with the somewhat narrow log-normal distribution employed in {\sc EddySed}.  We discuss the implications of a wider size distribution on transmission observations in Section \ref{sec:transmission}.

While the vertical extent of the clouds does not depend directly on $K_{zz}$, the assumption that $\left<v_\mathrm{sed}\right> = f_\mathrm{sed} w_\star$ determines the characteristic particle size at a location, and as a result, the characteristic particle size is a function of $K_{zz}$.  While {\sc EddySed} does include corrections for Reynolds number and Knutsen number (see \citealt{2001ApJ...556..872A}), in the limit where both of these quantities are small, the characteristic particle radius $r_\mathrm{g}$ becomes 
\begin{equation}
r_\mathrm{g} \propto \sqrt{\frac{f_\mathrm{sed}K_{zz}}{ L_\mathrm{mix}}} = \sqrt{\frac{K_{zz}}{L_\mathrm{cloud}}}\, .
\label{Eqn:Rg}
\end{equation}
\noindent  Through the dependence of $r_\mathrm{g}$ on $K_{zz}$ and its influence on the radiative properties of the clouds, $K_{zz}$ exerts an indirect influence on the atmospheric structure.

By default, {\sc EddySed} employs a parametrization of vertical convective mixing based on mixing-length theory from \citet{1985rapm.book..121G} (hereafter, GC85). In the GC85 mixing treatment, $w_\star$ is the convective mixing velocity and the eddy diffusion coefficient is taken to be
\begin{equation}
    K_{zz} = \max\left(\frac{H}{3}\left(\frac{L_\mathrm{mix}}{H}\right)^{4/3}\left(\frac{RF}{\mu\rho_a c_p}\right)^{1/3},K_{zz,\mathrm{min}}\right)\, ,
    \label{Eqn:KzzGC85}
\end{equation}
\noindent where $R$ is the universal gas constant, $F=\sigma T^4_\mathrm{eff}$ is the convective heat flux assuming an effective temperature $T_\mathrm{eff}=1130\,\mathrm{K}$\footnote{The value of $T_\mathrm{eff}$ has been chosen to retain consistency with \citet{lines_2019}; however, we note that they do not provide justification for the choice.}, $\mu = 2.33\, \mathrm{g\,mol^{-1}}$ is the mean molecular weight, $\rho_a$ is the atmospheric density, and $c_p = 1.3\times 10^4\,\mathrm{J\, K^{-1}}$ is the specific heat of the atmosphere at constant temperature.  The minimum value of $K_{zz,\mathrm{min}}= 10^5\,\mathrm{cm^2\,s^{-1}}$, taken from \citet{1981JGR....86.9707L},  is imposed to account for circulation-driven advection in radiative regions.  The associated mixing length $L_\mathrm{mix}$ is given by
\begin{equation}
    L_\mathrm{mix} = H\max(0.1,\Gamma/\Gamma_\mathrm{adiab}),
    \label{Eqn:LmixGC85}
\end{equation}
\noindent where $\Gamma$ and $\Gamma_\mathrm{adiab}$ are the local and dry adiabatic lapse rates.  
As the regions of interest within our target atmospheres are radiation-dominated (generally pressures below 10\,bar), vertical mixing is accomplished primarily though the large--scale flow, not convection, and while this is crudely accounted for through the inclusion of a minimum value on $K_{zz}$, studies of mixing in hot Jupiter atmospheres exist which may capture the mixing more accurately than the GC85 approach. \citet{2013A&A...558A..91P} (hereafter, P13) performed mixing studies using atmospheric tracers within a global climate model of HD209458b and fit the inferred mixing rate as a function of pressure ($P$),
\begin{equation}
K_\mathrm{zz} =\frac{5\times 10^4}{\sqrt{P/1\,\mathrm{bar}}}\, \mathrm{m^2s^{-1}}.    
\label{Eqn:KzzP13}
\end{equation}

\noindent  While this calculation of $K_{zz}$ is only strictly applicable to HD209458b, \citet{komacek_2019} have made analytic estimates of $K_{zz}$ for hot Jupiter atmospheres more generally which can be used when moving beyond HD209458b to other hot Jupiter atmospheres, and a discussion of how $K_{zz}$ estimate of \citet{komacek_2019} varies with planetary and stellar parameters can be found in \citet{baeyens_2021}. 

In this study, we implement the P13 mixing treatment, but retain the same minimum $K_{zz,\mathrm{min}}$ applied as above, although within the model atmosphere we consider, the minimum value is never reached. P13 consider their parametrization to be valid between pressures of a few bar and a few $\mu$bar.  While in many of our simulations this range of pressures encompasses where clouds are found,   the cloud deck may be at higher pressures and we remain cognizant that our mixing treatments may not be accurate.
We additionally assume that the mixing length is equal to the pressure scale height, $L_\mathrm{mix} = H$.  A similar choice was used in \citet{2017A&A...600A..10M}.  The results are ultimately insensitive to rescaling $L_\mathrm{mix}$ by a constant factor as $L_\mathrm{mix}$ only influences the physics through $L_\mathrm{cloud} = f_\mathrm{sed}^{-1}L_\mathrm{mix}$ and so any constant rescaling of $L_\mathrm{mix}$ can be offset by a rescaling of $f_\mathrm{sed}$.  Due to the more complicated functional form of the mixing length in the GC85 model (equation \ref{Eqn:LmixGC85}), a constant rescaling to move between models is not possible; however, it can be done in an approximate fashion, as discussed below.

The differing assumptions about $L_\mathrm{mix}$ between our GC85 and P13 models directly impacts the interpretation of the results, as atmospheres with identical sedimentation factors will have clouds with differing vertical extents.  In the case of GC85, the mixing scale is a function of the atmospheric properties and as such isn't a constant fraction of H.  Averaging over the either GC85 simulation volume, we find that $\left<L_\mathrm{mix}/H\right>_V$ is $0.2$, suggesting that simulations with similar cloud structures may differ by a factor of 5 in $f_\mathrm{sed}$ when comparing between mixing treatments.

In either mixing treatment, we view $K_{zz}$ as approximating relevant transport in the GCM regardless of scale without the need for more computationally intensive schemes. The switch to a mixing model based on a global circulation does, however, raise issues of applicability. This is as {\sc EddySed}, when adopting the convective mixing treatment of GC85, was intended to balance the local upward mixing against the local downward sedimentation.  In the P13 model, however, it is global circulation that is responsible for maintaining the mixing ratios of condensable species at higher altitudes, and the assumption that the global upward mixing locally balances the downward sedimentation may be inappropriate. One possible solution is instead to balance the local upward bulk motions with sedimentation; however, in cases where the vertical velocities are small or negative, solutions become problematic or non-existent.  Addressing this issue likely involves moving beyond {\sc EddySed} and will be discussed in a future paper.  For this work, we are primarily interested in exploring the impact of changing the mixing treatment in a widely used cloud model.

To couple {\sc EddySed} to the UM, vertical columns are passed to {\sc EddySed} independently at each radiation time step, with the returned cloud radiative properties (opacity, single-scattering albedo, and asymmetry parameter) for each wavelength bin passed to the radiation solver. The resulting heating rates are then used by the rest of the model to evolve the atmospheric structure before the process repeats. For each cloud species, the average radiative properties are calculated using the log-normal size distribution outlined above and a table of pre-calculated Mie coefficients.  The table has been expanded from \cite{lines_2019} and now includes 54 radius bins between $10^{-7}\, \mathrm{cm}$ and $0.11\,\mathrm{cm}$.  While it does add computational expense to average optical properties across the size distribution as opposed to simply using a characteristic particle size, using a single characteristic particle size can impact both the size and shape of the transmission spectra in unpredictable ways \citep{Powell_2019}. 

To ensure numerical stability, we modify the cloud opacities in two ways.  First, cloud opacities are slowly ramped to their full value over the first 100 days. This avoids large heating rates associated with the cloudless initial state being a large departure from the final cloudy state.  A similar ramping was employed in \citet{lines_2019}.  Second, we include a linear opacity limiter above $1\, \mathrm{mbar}$.  For runs with cloud scales larger than the pressure scale height, specifically the P13 simulation with $f_\mathrm{sed} = 0.1$, the relatively large cloud mass at low pressures combined with the small particle radii required to suspend clouds at those pressures results in significant opacity and the potential for numerical instability.

In the context of this study, there are a number of limitations to our approach. As {\sc EddySed} models a single column in equilibrium each time it is called, there is no possibility for horizontal advection of condensate or time evolution of the cloud within our simulations. Moreover, by employing a globally-averaged mixing rate, the model only captures local mixing to the extent that local mixing influences the average. There is no variation in the mixing with latitude  thus potentially limiting variation between the poles and the equatorial region. As {\sc EddySed} also ignores the microphysics of particle growth and evaporation, it may be the case that the chemical timescales are comparable to or shorter than the timescale for horizontal advection and a globally-averaged mixing treatment may no longer be appropriate (\citealt{2018ApJ...866....2Z}; see \citealt{2019AJ....158..244C} for a model which attempts to address the issue of the growth timescale).

\subsection{Simulations}
\label{subsec:simulations}
To investigate the impact of the choice of mixing treatment we perform a suite of simulations of HD209458b similar to those of \citet{lines_2019}, using identical 3D initial conditions to aid comparability. The choice to model HD209458b is done for consistency with \citet{lines_2019} and because HD209458b serves as a benchmark in the field. In \citet{lines_2019}, two separate 3D initial conditions were investigated: a standard deep interior model (SDI) and a hot deep interior model (HDI), neither of which include clouds.  The SDI initial condition was created using a 1D pressure-temperature profile generated by the radiative-convective code ATMO \citep{2015ApJ...804L..17T,drummond_2016} with a 3D simulation subsequently evolved for 1200 days, reaching a quasi-steady state.  The HDI initial condition is generated similarly except that the initial 1D temperature profile is increased by $800\, \mathrm{K}$ in each layer before the 3D simulation is evolved for 800 days. Note that although the total elapsed simulation times for these two setups differ, both have reached a state of steady wind velocities in the upper atmosphere. In particular, as the higher pressure regions of standard setup hot Jupiter simulations gradually warm \citep[see][for example]{2016A&A...595A..36A}, the hotter interior profile is likely closer to the final state \citep[see][]{sainsbury_martinez_2019}. A discussion of these initial conditions can be found in \citet{lines_2019}, with the latter motivated by the warming of the deep atmosphere in hot Jupiters simulations \citep[see][for details]{2016A&A...595A..36A,tremblin_2017,sainsbury_martinez_2019}.   In this work, we focus on the HDI case as \citet{lines_2019} found it to have reasonable agreement with observations, and our intention is to only investigate the impact of changing the mixing treatment.  In addition to serving as the 3D initial conditions for the cloudy simulations performed in this paper, the HDI initial condition also serves as the ``Clear Sky'' case used for comparison with the cloudy results to isolate the impact of the clouds on the atmospheric structure.

Given the initial conditions, we perform a suite of simulations varying $f_\text{sed}$ using both the GC85 mixing treatment as well as the P13 mixing treatment. The value for $f_\mathrm{sed}$ is only minimally constrained in models using the GC85 mixing treatment, with a range of values extending from $f_\mathrm{sed} \sim 1-10$ in brown dwarf atmospheres \citep{saumon_2008} to $f_\mathrm{sed} \sim 0.01$ in cloudy super-Earths \citep{morley_2015}.  These values are set by observations instead of being motivated by the underlying physics.  A study by \citet{gao_2018} which attempted to calibrate $f_\mathrm{sed}$ in brown dwarf atmospheres using the {\sc CARMA} microphysical cloud model \citep{turco_1979,toon_1979} found $f_\mathrm{sed} \sim 0.1$ with the value sensitive to nucleation rates and other material properties and, to a lesser extent, $K_{zz}$. Lacking physical motivation for a specific choice of $f_\mathrm{sed}$ in hot Jupiter atmospheres, we perform simulations across a range of values of $f_\mathrm{sed}$ with the intention to cover a broad range of cloud scales.  As in \citet{lines_2019}, we investigate both the $f_\mathrm{sed}=0.1$ and $f_\mathrm{sed}=1.0$ cases.   Given how simulations with differing values of $f_\mathrm{sed}$ can have similar cloud scales $L_\mathrm{cloud}$ when comparing between mixing models, we additionally perform simulations with $f_\mathrm{sed}=2.0$ and $5.0$ for the P13 case.

\section{Results}
\label{sec:results}

In this section we present the thermodynamic and dynamic structures resulting from our simulations, before detailing the mixing and cloud distributions themselves, and finishing with an analysis of the observational implications. 

 \subsection{Atmospheric Structure}
 \label{subsec:atm_struct}
 
% We begin with a direct comparison of our simulations with the results of \citet{lines_2019} for the cases of $f_\mathrm{sed} = 0.1$ and $1.0$.
 
\begin{figure*}
	\includegraphics[width=\textwidth]{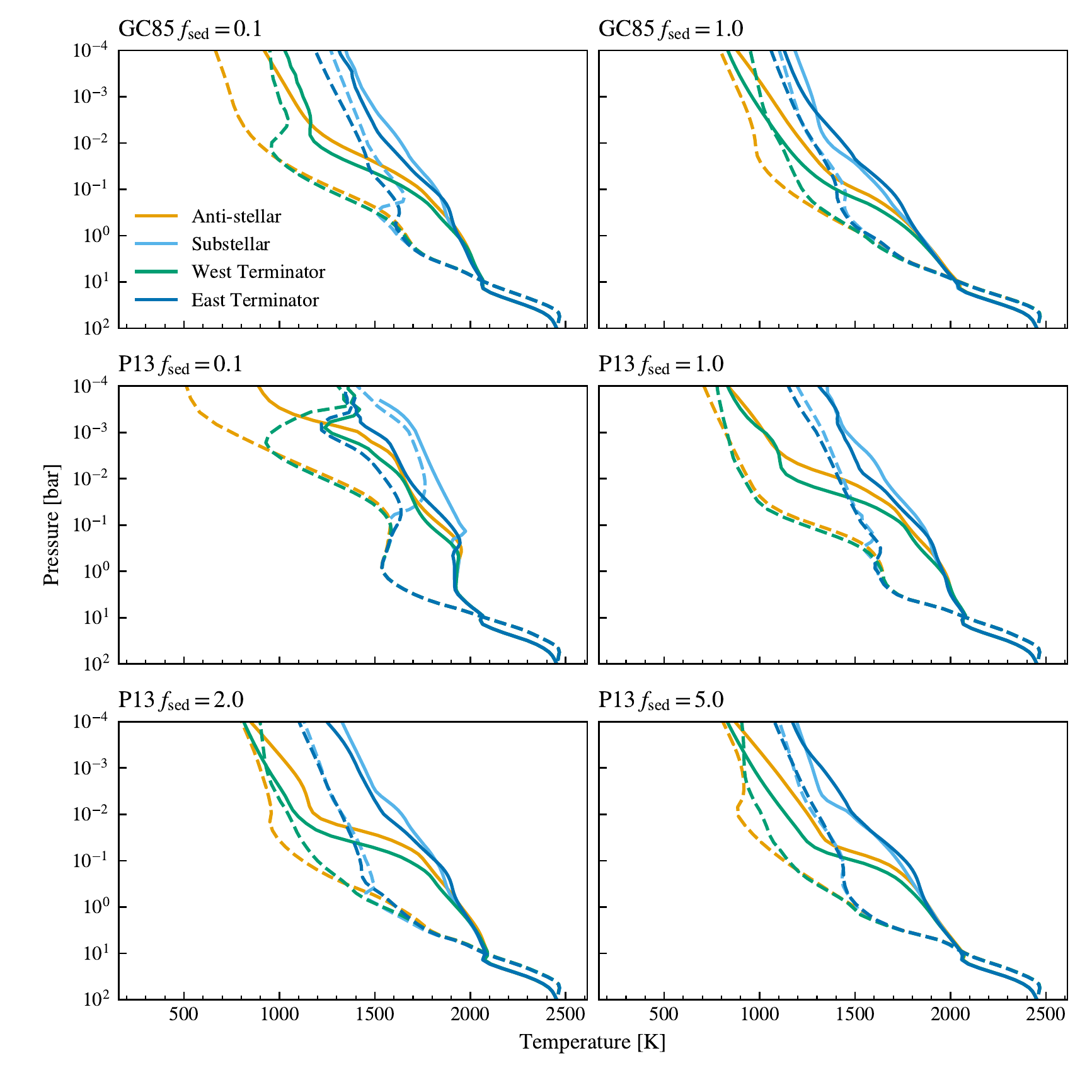}
    \caption{Pressure-temperature profiles for each simulation outlined in Section \ref{subsec:simulations}. The equatorial profiles (solid lines) and mid-latitude profiles (taken at a latitude of $45\degree$; dashed lines) at the substellar point, antistellar point, and eastern and western terminators are shown. The similarities between the GC85 $f_\mathrm{sed}=1.0$ and P13 $f_\mathrm{sed}=5.0$ cases are due in part to the similar cloud scale heights, as discussed in section \ref{subsec:eddysed}.  }
    \label{Fig:PT}
\end{figure*}

\begin{figure*}
	\includegraphics[width=\textwidth]{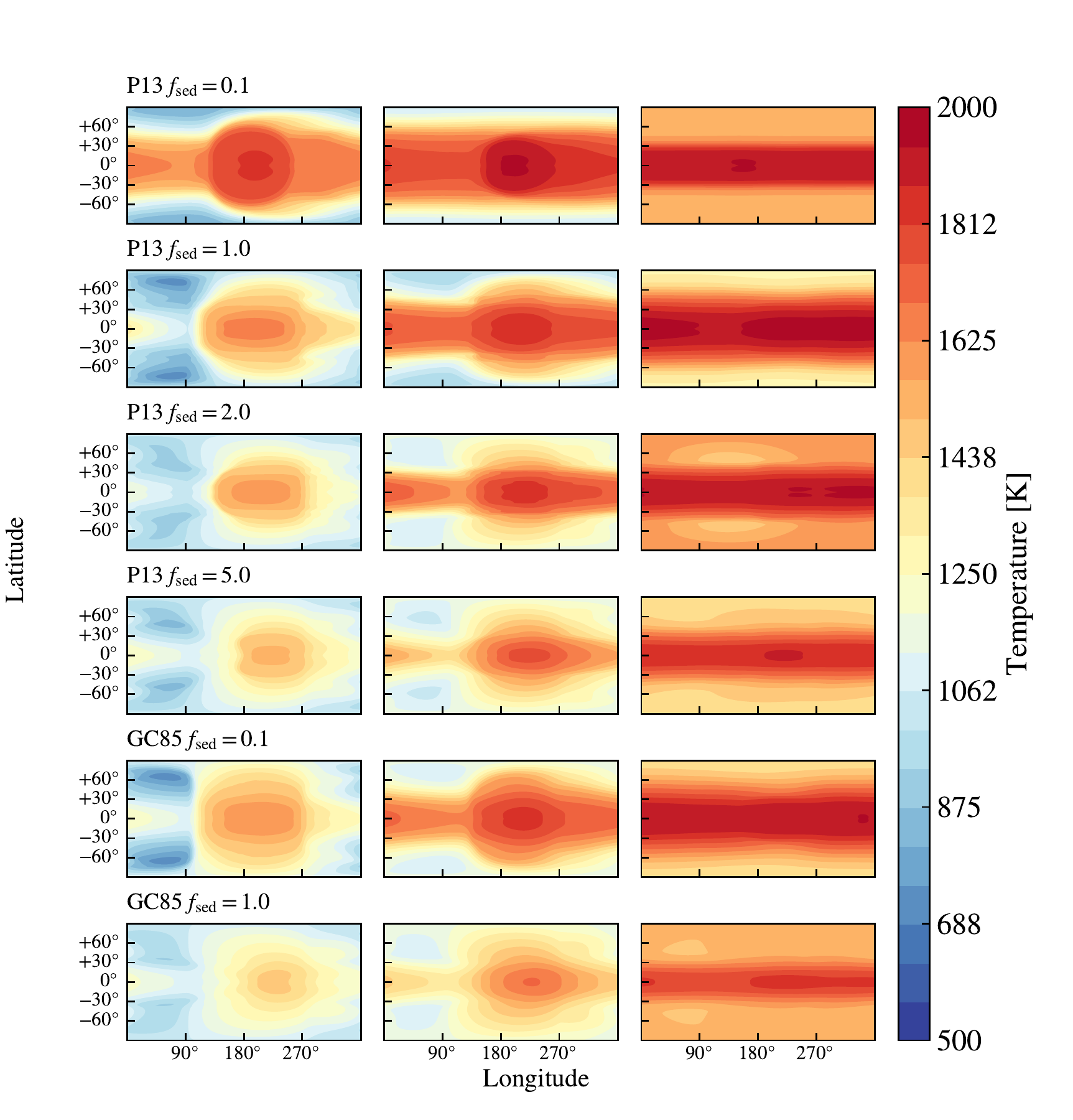}
    \caption{Temperature at 0.01 bar (left column), 0.1 bar (middle column) and 1 bar (right column) for each simulation described in Section \ref{subsec:simulations}. The substellar point is located at the center of each panel (i.e., at a longitude of $180\degree$).  The top four rows correspond to the P13 simulations while the two bottom rows correspond to the GC85 simulations. For runs with the same mixing treatment, the lower values of $f_\mathrm{sed}$ correspond to increasing cloud vertical extent. The similarities between the GC85 $f_\mathrm{sed}=1.0$ and P13 $f_\mathrm{sed}=5.0$ cases are due in part to the similar cloud scale heights, as discussed in section \ref{subsec:eddysed}.  }
    \label{Fig:TempPanel}
\end{figure*}

\begin{figure*}
	\includegraphics[]{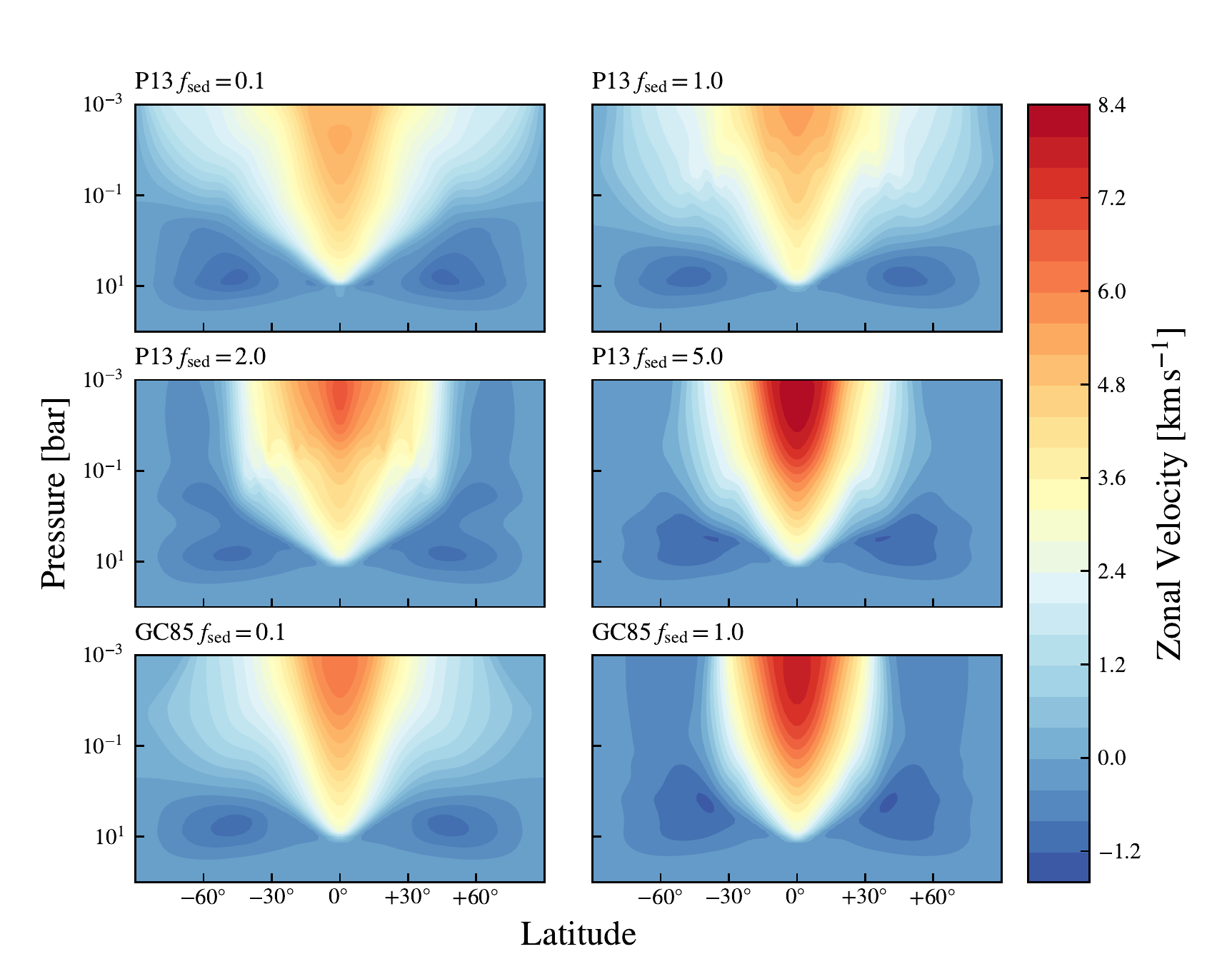}
    \caption{The azimuthally-averaged zonal velocity.  The equatorial jet is clearly visible and both mixing treatments show a trend of reduced peak zonal velocity with cloud extent. The similarities between the GC85 $f_\mathrm{sed}=1.0$ and P13 $f_\mathrm{sed}=5.0$ cases are due in part to the similar cloud scale heights, as discussed in section \ref{subsec:eddysed}.  }
    \label{Fig:ZonalVel}
\end{figure*}

We begin with a direct comparison of our P13 simulations with the HDI GC85 simulations performed in \citet{lines_2019} (i.e., our GC85 $f_\mathrm{sed} = 0.1$ and $1.0$ cases). All simulations exhibit the characteristic hot zonal jet with a dayside hotspot observed at pressures less than 1 bar (see Figs \ref{Fig:PT} and \ref{Fig:TempPanel}). Along the equator, we see similar temperature profiles at the eastern terminator as at the substellar point due to warm gas being advected by the equatorial jet.  Similarly, the temperature profiles at the western terminator resemble those at the anti-stellar points. The P13 simulations show much more pronounced hotspots with higher dayside peak temperatures compared to the GC85 simulations with the same sedimentation factor.  The hotspots themselves are shifted eastward for more compact clouds (i.e., increasing $f_\mathrm{sed}$).  The GC85 simulations exhibit a greater hotspot offset compared to the P13 simulations, both when comparing simulations with identical values of $f_\mathrm{sed}$ and when accounting for a factor of 5 difference in $\left<L_\mathrm{mix}\right>_V$. The GC85 simulations also exhibit somewhat faster zonal winds with the $f_\mathrm{sed} = 1.0$ case having a peak zonally-averaged speed of $7.7\,\mathrm{km\,s^{-1}}$ compared to $5.5\, \mathrm{km\,s^{-1}}$ for the P13 case with the same value of $f_\mathrm{sed}$ (see Fig. \ref{Fig:ZonalVel}).  In general, smaller values of $f_\mathrm{sed}$ correspond to slower zonal winds, although the effect is less pronounced in the P13 runs. 

As discussed in Section \ref{subsec:eddysed}, the cloud scales $L_\mathrm{cloud}$ differ on average by roughly a factor of 5 between GC85 and P13 which results in the clouds in P13 models having significantly greater vertical extent for the same value of $f_\mathrm{sed}$.  Therefore, it is more appropriate to compare GC85 $f_\mathrm{sed}=1.0$ with P13 $f_\mathrm{sed}=5.0$ as they have similar, although not identical cloud scale heights. They exhibit similar atmospheric temperatures (Fig. \ref{Fig:TempPanel}, fourth and sixth rows) and the zonal wind peaks at $7.8\,\mathrm{km\,s^{-1}}$ in the GC85 simulation compared to $8.3\,\mathrm{km\,s^{-1}}$ in the P13 simulation.    The differences between the simulations are due to cloud scale heights agreeing only as an atmospheric average.  For the GC85 case, $L_\mathrm{cloud}$ exhibits local variation due to its temperature dependence, which is not something that occurs within the P13 simulation where $L_\mathrm{cloud} = f_\mathrm{sed}^{-1}H$.  The differing treatments for $K_\mathrm{zz}$ also impacts the thermal structure due to the dependence of the particle radius and thus cloud opacities on the atmospheric mixing.

All the temperature profiles exhibit a temperature offset at high pressures ($P \gtrsim 10\,\mathrm{bar}$) between mid-latitudes and the equator. This is seen in both clear and cloudy models and is thus unrelated to the cloud model being investigated; however, as the models are not converged at these pressures (see \citealt{tremblin_2017,mayne_2017}), it is unclear whether this effect will be seen in a converged solution or if it is simply a transient feature.

\subsection{Mixing and Cloud Structure}
\label{subsec:clouds}

\begin{figure}
	\includegraphics[width=\columnwidth]{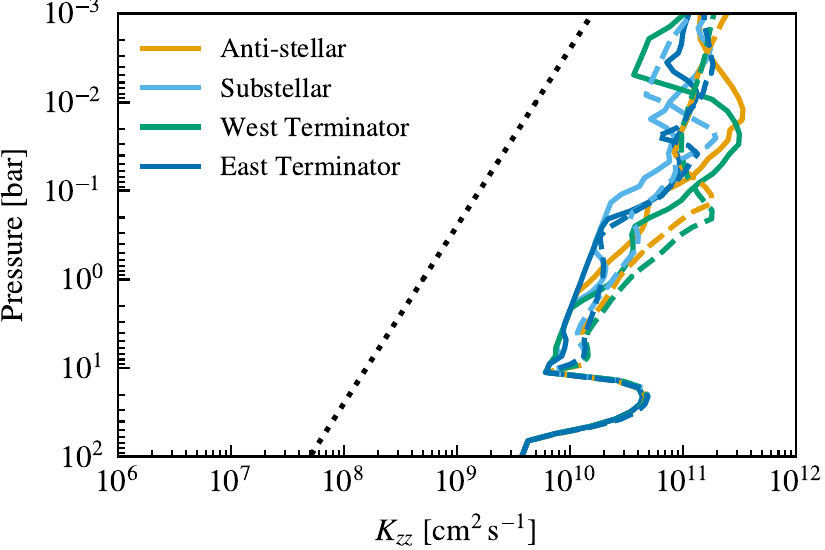}
    \caption{The equatorial profiles of $K_{zz}$ at four longitudes for the GC85 $f_\mathrm{sed} = 0.1$ (solid lines) and $f_\mathrm{sed} = 1.0$ (dashed lines) cases.  The $K_{zz}$ for the P13 model is shown as a dotted black line.}
    \label{Fig:Kzz}
\end{figure}

The equatorial profiles of $K_{zz}$ for the GC85 and P13 simulations are shown in Fig. \ref{Fig:Kzz}, and, in general, the GC85 mixing rates are approximately one to two orders of magnitude larger than the P13 mixing rates.  As $K_{zz}$ in the GC85 case depends on atmospheric properties beyond the local pressure (see equation \ref{Eqn:KzzGC85}), there is an order of magnitude variation in mixing with longitude.   The P13 simulations, by contrast, shows no variation beyond the pressure dependence (see Fig. \ref{Fig:Kzz}, dotted line); however, this is a consequence of equation \ref{Eqn:KzzP13} being a fit to the zonal average mixing and not necessarily an indication that the underlying tracer model lacked such a dependence.

\begin{figure}
	\includegraphics[width=\columnwidth]{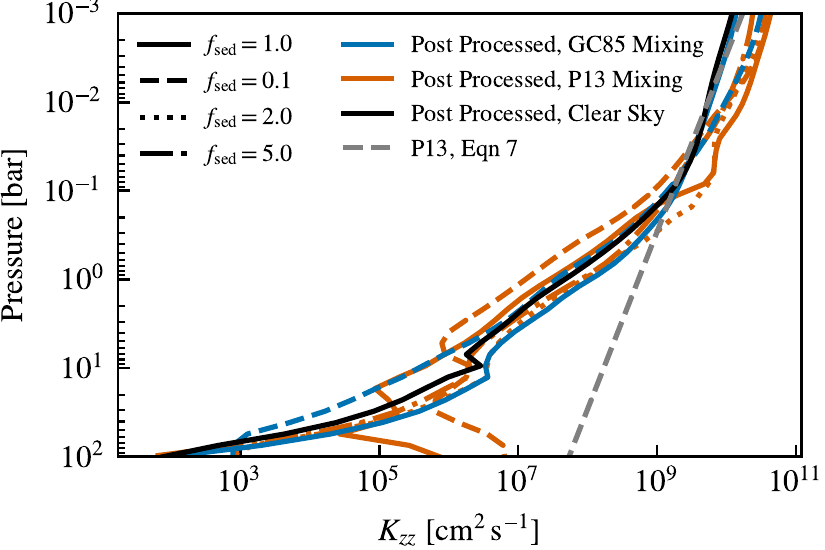}
    \caption{The estimated $K_{zz}$ profiles from equation  \ref{Eqn:KzzKomacek} for each simulation. The P13 $K_{zz}$ (equation \ref{Eqn:KzzP13}) is in grey for comparison. }
    \label{Fig:KzzEstimate}
\end{figure}

The $K_{zz}$ used in the P13 mixing treatment (equation \ref{Eqn:KzzP13}) is based on clear sky simulations and changes to the atmosphere caused by the presence of clouds may not be accounted for.  This may be especially important in the P13 $f_\mathrm{sed} = 0.1$ simulation where we see the largest changes in temperatures as well as the presence of dayside temperature inversions.  To understand the degree to which mixing rates might be impacted, we estimate $K_{zz}$ for each simulation, using the analytic estimate of \citet{komacek_2019},

\begin{equation}
K_{zz} \sim \frac{\overline{w}^2}{\tau_\mathrm{adv}^{-1} + \tau_\mathrm{chem}^{-1}} \sim \overline{w}^2 \tau_\mathrm{adv}\,\, ,
\label{Eqn:KzzKomacek}
\end{equation}

\noindent where $\overline{w}$ is the root-mean-squared  of the vertical speed along isobaric pressure contours and $\tau_\mathrm{adv} = R_\mathrm{p}/\overline{u}$ is the horizontal advection timescale with the horizontal wind speed $\overline{u}$ similarly averaged along isobaric pressure contours.  We ignore the chemical timescale $\tau_\mathrm{chem}$ in our analysis, effectively taking it to be infinitely long, for consistency with the tracer study of P13; however, it likely becomes important in regions where the horizontal velocities are small such as at high pressures.  The resulting mixing rates are shown in Fig. \ref{Fig:KzzEstimate}. For $P \lesssim 10\,\mathrm{bar}$, we see roughly an order of magnitude variation between simulations; however, there is no obvious trend with $f_\mathrm{sed}$ or with mixing treatment.   At higher pressures ($P \gtrsim 10\,\mathrm{bar}$), we see large variation between simulations, as well as larger fluctuations with pressure within simulations.  At these pressures, however, the simulations are not converged (see \citealt{tremblin_2017,mayne_2017}) and running the simulations to convergence may reduce the variations between simulations as well as the fluctuations in time.  The inclusion of a finite chemical timescale will also impact the estimated $K_{zz}$ as $\tau_\mathrm{adv}$ becomes very large at these pressures.   As the variation between simulations in the estimated $K_{zz}$ is limited at pressures where we find clouds, we take this as evidence for limited impact of clouds on the overall mixing.  Comparing with the P13 $K_{zz}$, we find agreement to within an order of magnitude at low pressures ($P \lesssim 1\,\mathrm{bar}$); however, the analytic estimates diverge at higher pressures.   As P13 only claim their fit to be valid for $P \lesssim 1\,\mathrm{bar}$, this divergence is not surprising.

% Transition - Kzz doesn't impact distribution of condensate directly.

\begin{figure*}
	\includegraphics[width=\textwidth]{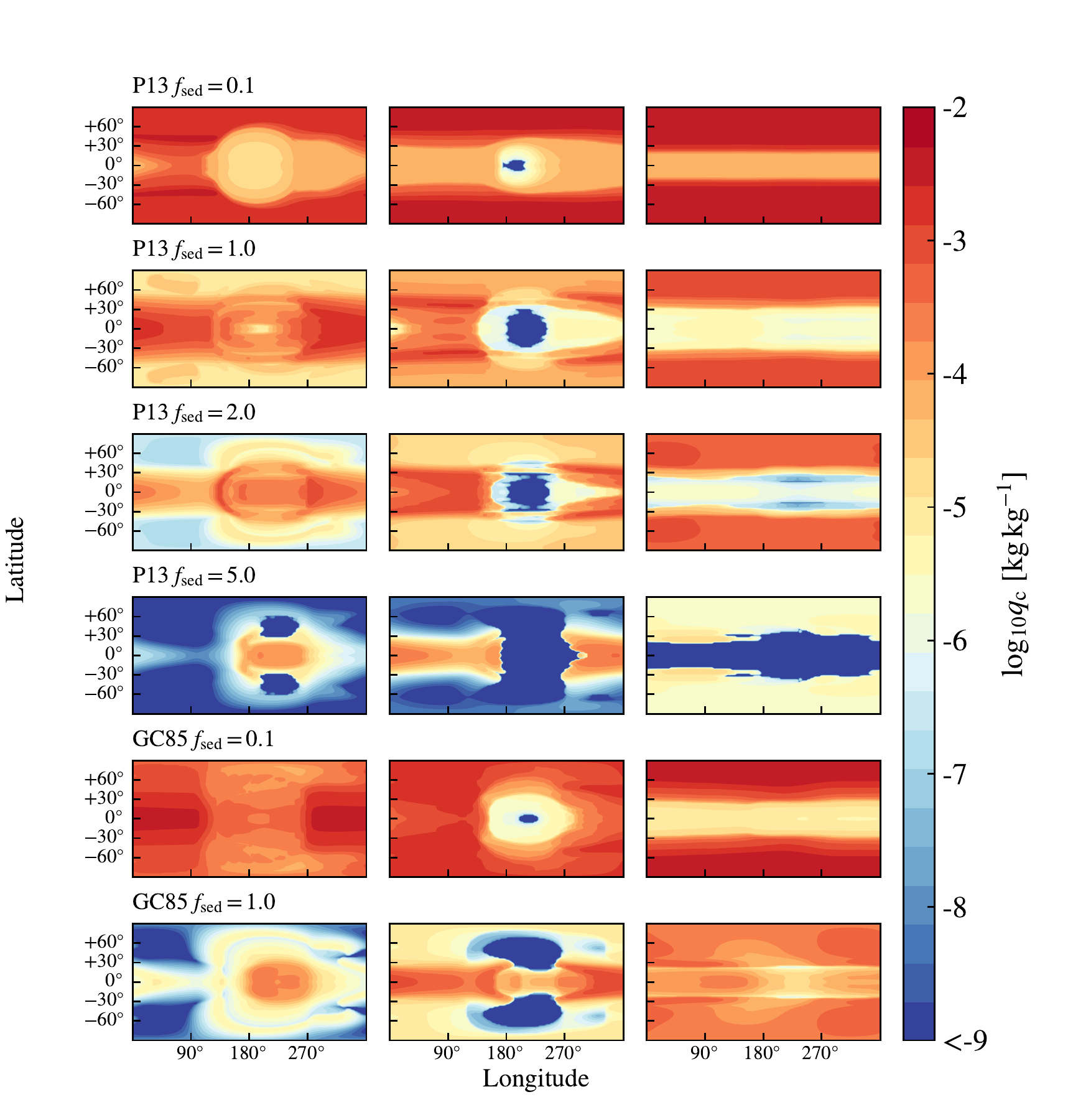}
    \caption{The total condensate mixing ratio at 0.01 bar (left column), 0.1 bar (middle column) and 1 bar (right column) for each simulation described in Section \ref{subsec:simulations}. The substellar point is located at the center of each panel (i.e., at a longitude of $180\degree$).  The top four rows correspond to the P13 simulations while the two bottom rows correspond to the GC85 simulations.  The observed ring-like features around the substellar point are due to contributions from cloud species with differing critical temperatures. }
    \label{Fig:QcPanel}
\end{figure*}

The distribution of condensates exhibit a strong dependence on latitude as well as a day-night asymmetry.  The hot equatorial jet shows reduced condensate mixing ratios at higher pressures ($\sim 10^2$ bar) in all the simulations with only $\mathrm{Al_2O_3}$ near the equator  (see Fig. \ref{Fig:QcPanel}, right column) with the cooler poles showing the highest mixing ratios of condensate, primarily consisting of $\mathrm{Al_2O_3}$, $\mathrm{MnS}$, $\mathrm{MgSiO_3}$, $\mathrm{Mg_2SiO_4}$, $\mathrm{Fe}$, and $\mathrm{Cr}$.  Moving to lower pressures, the day-night asymmetry becomes apparent, with the day-side hot spot remaining free of condensate except for $\mathrm{Al_2O_3}$ and the night-side exhibiting clouds around the equator.  At the poles, the behaviour depends on the efficiency of sedimentation (i.e., on $f_\mathrm{sed}$).  As clouds form at relatively high pressures near the poles, sedimentation reduces the vapour mixing ratio at lower pressures above the clouds, reducing further cloud formation.   As a result, we see the greatest amount of condensate at the poles in the models with $f_\mathrm{sed} = 0.1$ where the effects of sedimentation are suppressed. (see Fig. \ref{Fig:QcPanel}, first and fifth rows). 

\begin{figure*}
	\includegraphics[width=\textwidth]{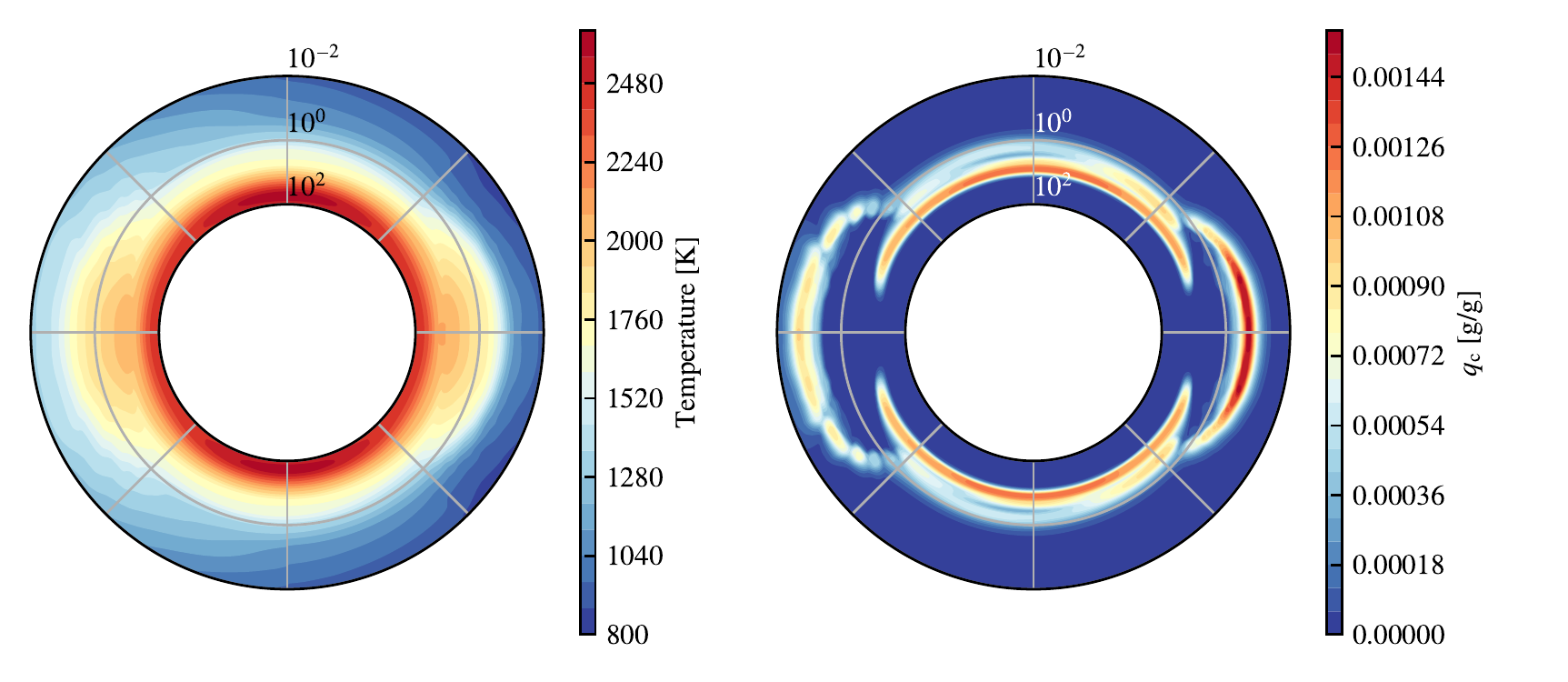}
    \caption{A slice through the terminator for the P13 $f_\mathrm{sed}=2.0$ model showing temperature (left) and total condensate mixing ratio (right).  The north pole is at the top of each plot with the radial direction corresponding to pressure, in bars. The effect of the hot equatorial jet pushing the cloud deck to lower pressures can be seen.}
    \label{Fig:Terminator}
\end{figure*}

This impact on cloud morphology can be seen by examining the distribution at the terminator.  Fig. \ref{Fig:Terminator} shows slices through the terminator of the P13 $f_\mathrm{sed} = 2.0$ simulation, chosen as it exhibits sufficient cloud opacity to impact observations.   The peak in total condensates near the poles can be seen to occur at $P\sim 10\,\mathrm{bar}$.  At the equator, the higher temperatures inhibit cloud formation, pushing the cloud deck to $P\sim 0.1\,\mathrm{bar}$, ultimately forming `arches' above the hottest regions of the equatorial jet.   As the eastern terminator (the left side of each plot in Fig. \ref{Fig:Terminator}) is hotter (left panel), we see the cloud deck pushed to lower pressures on the eastern terminator, resulting in the eastern limb having a larger effective radius (see Section \ref{subsec:observations}).

%We do not find significant $\mathrm{Na_2S}$ or  $\mathrm{ZnS}$ clouds except for at low pressure ($P \lesssim 10 \,\mathrm{mbar}$) on the night side near the poles.  It is unclear whether these clouds would form due to the large nucleation energy barriers associated with sulphides \citep{gao_2018, Powell_2019}.   $\mathrm{MnS}$ and $\mathrm{Fe}$ have considerable cloud coverage in our simulations and have similar uncertainties regarding their ability to realistically condense out.  \citet{Powell_2019} found that $\mathrm{Fe}$ was able to form clouds despite the energy barrier due to large supersaturations.  

There is uncertainty as to whether sulphides and iron are capable of condensing to form clouds due to large energy barriers.  While we do not see significant $\mathrm{Na_2S}$ or  $\mathrm{ZnS}$ cloud coverage except in limited regions at lower pressures ($P \lesssim 10 \,\mathrm{mbar}$), we do see significant $\mathrm{MnS}$ and $\mathrm{Fe}$ cloud coverage in our simulations.  In the microphysical simulations of \citet{Powell_2019}, $\mathrm{MnS}$ clouds did not form but some $\mathrm{Fe}$ clouds were observed despite the energy barrier due to large supersaturations.  Due to the possibility of these clouds influencing the atmospheric evolution, future simulations excluding these species might be necessary.

% Average particle size

\begin{figure*}
	\includegraphics[width=\textwidth]{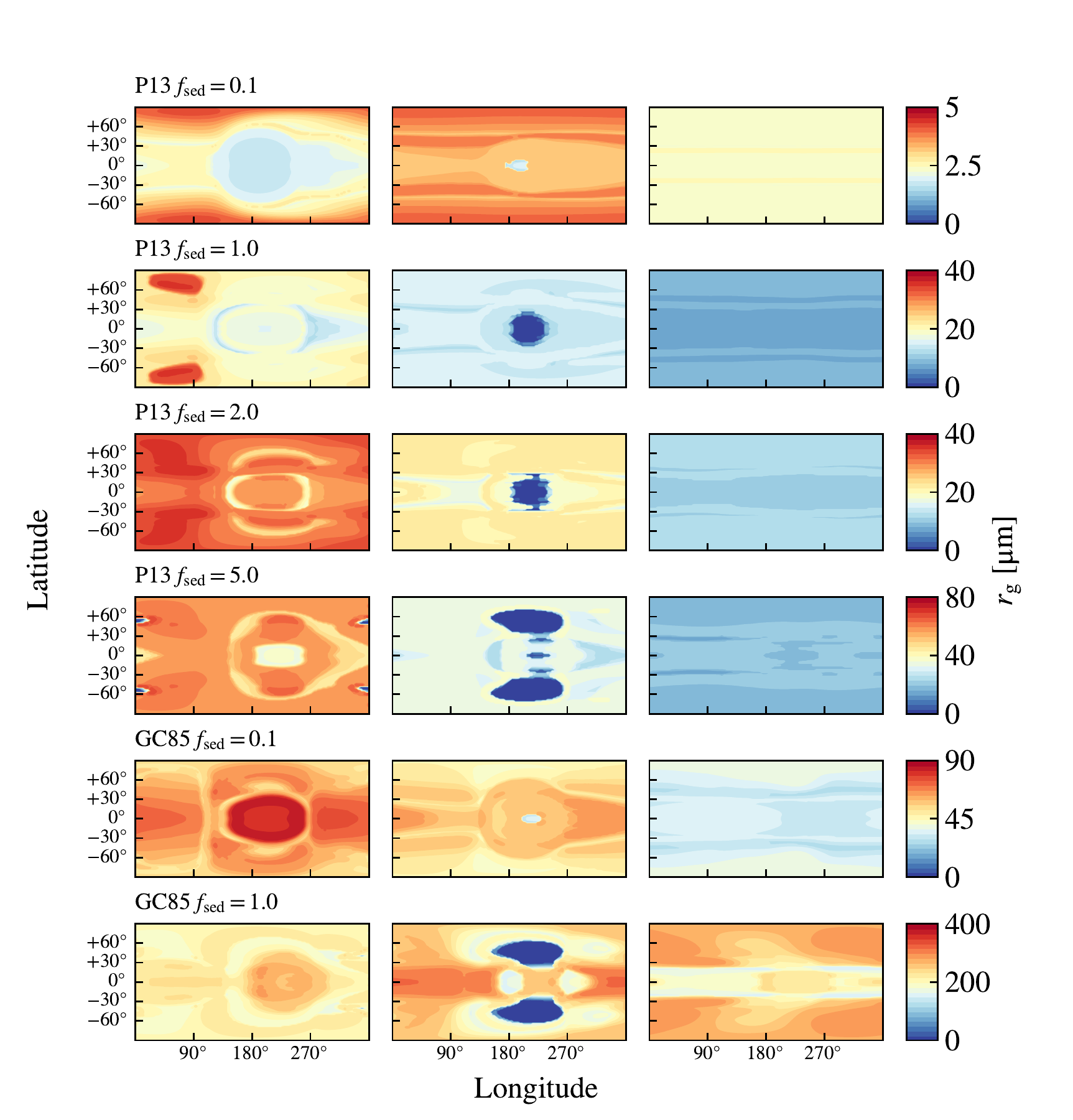}
    \caption{The characteristic cloud particle radius $r_\mathrm{g}$ at 0.01 bar (left column), 0.1 bar (middle column) and 1 bar (right column) for each simulation described in Section \ref{subsec:simulations}. The substellar point is located at the center of each panel (i.e., at a longitude of $180\degree$).  The top four rows correspond to the P13 simulations while the two bottom rows correspond to the GC85 simulations.}
    \label{Fig:RgPanel}
\end{figure*}

As the model contains clouds of multiple species, we calculate the mass-mixing-ratio-weighted particle size throughout the atmosphere,

\begin{equation}
    \left<r_\mathrm{g}\right> = \frac{\sum_s r_{\mathrm{g},s} q_s}{\sum_s q_s} \,\,,
\end{equation}

\noindent where $r_{\mathrm{g},s}$ is the cloud particle radius for species $s$ and $q_s$ is the associated mass mixing ratio.   Fig.\ref{Fig:RgPanel} shows $\left<r_\mathrm{g}\right>$ for each  simulation.  At higher pressures ($\sim 1\, \mathrm{bar}$, Fig. \ref{Fig:RgPanel}, right column), the simulations show relatively homogeneous particle size distributions with some latitudinal variation between the equatorial jet and the poles.  As pressures decrease, however, the structures become more diverse, with a general trend of larger particle sizes at lower pressures.  This is due to the direct dependence of  the particle size on $K_{zz}$ (see equation \ref{Eqn:Rg}).  As is seen in \citet{lines_2019}, this trend continues until $P \sim 0.01\,\mathrm{bar}$, where particle sizes begin to decrease with lower pressures due to the reduction in atmospheric drag. 

% Discussion of cloud types

 \subsection{Observations}
  \label{subsec:observations}
  
  In this section, we compare synthetic observations derived from our simulations to each other and to observational data, including transmission (Section \ref{sec:transmission}) and emission spectra (Section \ref{sec:emission}) and phase curves (Section \ref{sec:phase}).
  
  \subsubsection{Transmission}
  \label{sec:transmission}
 \begin{figure*}
	\includegraphics[]{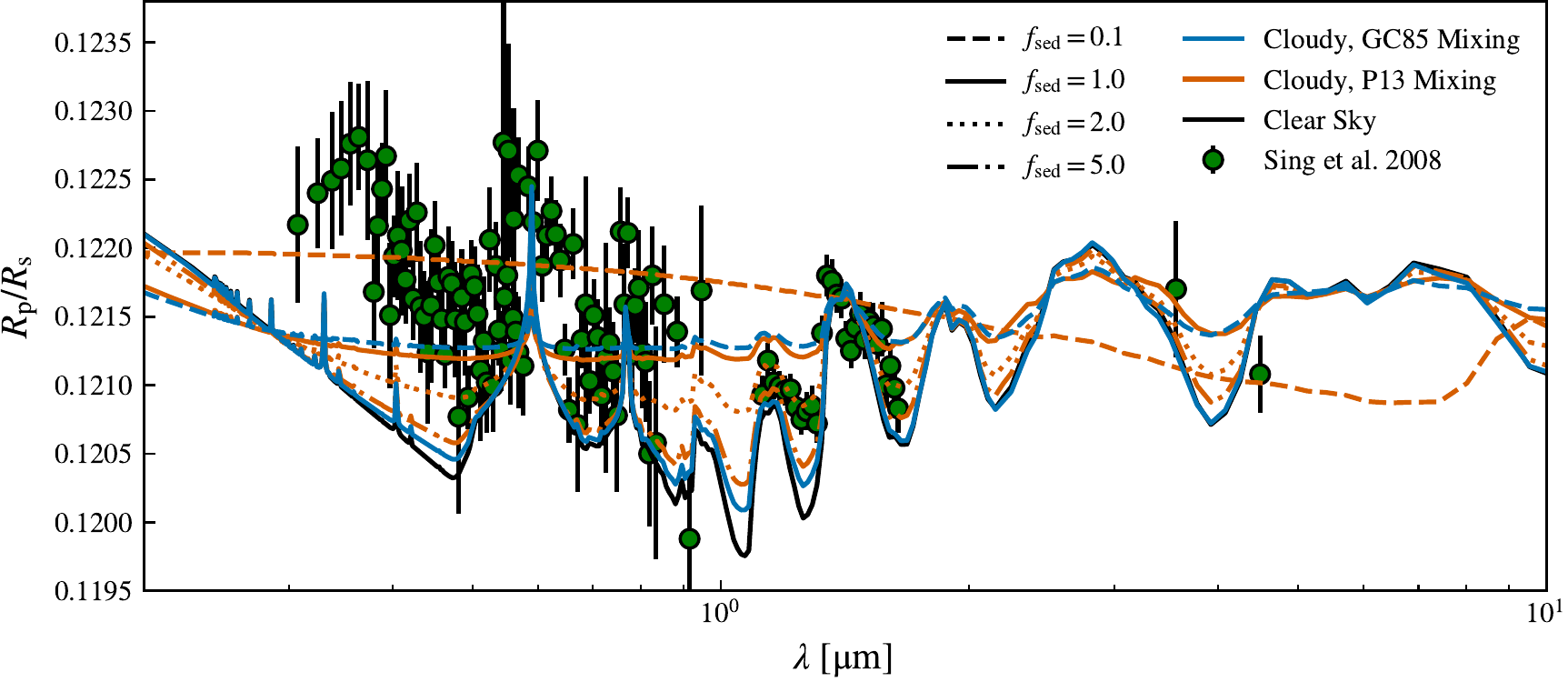}
    \caption{Transmission spectra for all simulations with GC85 mixing (blue lines) and P13 mixing (red lines). The clear sky spectra is in black. Observations from \citet{2008ApJ...686..667S} are shown in green. }
    \label{Fig:Transit}
\end{figure*}

For each simulation, we generated synthetic transmission spectra using the 3D transmission scheme described in \citet{2018MNRAS.481..194L} and employed previously in \citet{lines_2019} with the resulting spectra converted to an effective radius relative to the stellar radius.  The method used in \citet{2018MNRAS.481..194L} only includes the contribution from one side of the planet, either the dayside or the nightside, and doubles the result and thus does not capture the effects of gradients across the terminator.  To address this, \citet{lines_2019} computed the transits for the dayside and the nightside  and used the simple mean along each line of sight to compute a final result.  Here we instead take the geometric mean of the transits which has the effect of summing the underlying dayside and nightside optical depths.  This modification results in only minor differences in the transmission spectra. 

Fig. \ref{Fig:Transit} shows the transmission spectra for all the simulated atmospheres as well as the observed transmission spectrum from \citet{2008ApJ...686..667S}.  As the pressure of the atmosphere at the measured planetary radii is not precisely known,, all the model spectra have been scaled so that they agree with the observations at $1.4\,\mathrm{\mu m}$.  The clear sky spectrum is shown in black and exhibits prominent sodium, potassium, and water features, which are also seen in the observed spectrum.  By comparison, many of the cloudy simulations exhibit flat spectra with muted spectral features. This is a direct consequence of those simulations having clouds with large vertical extents (i.e., large $L_\mathrm{cloud}$).  A naive comparison of the GC85 and P13 simulations with $f_\mathrm{sed}=1.0$ show differing results, with the GC85 simulation resembling the the clear spectrum and the P13 simulation exhibiting a much flatter, cloudier spectrum.  These differences are due primarily to differences between the mixing lengths  and how the sedimentation factor is defined rather than the efficiency of sedimentation itself. Comparing instead the GC85 simulation with $f_\mathrm{sed} = 1.0$ and the P13 simulation with $f_\mathrm{sed} = 5.0$ which both have $\left<L_\mathrm{cloud}\right> \sim 0.2H$, we see similar spectra.   The differences can be attributed to the impact of particle size and local variation in the mixing length, as discussed above.

As both $f_\mathrm{sed}=0.1$ and $f_\mathrm{sed} = 1.0$ P13 simulations exhibit relatively flat spectra due to their cloud content, we include additional simulations with $f_\mathrm{sed}=2.0$ and $f_\mathrm{sed} = 5.0$ (i.e., $L_\mathrm{cloud} = 0.5H$ and $0.2H$, respectively; see Fig.\ref{Fig:Transit}, dotted and dash-dotted red lines).  These simulations, especially the $f_\mathrm{sed}=2.0$ case,  show better agreement with the observations in the $1-2\,\mathrm{\mu m}$ part of the spectrum by allowing cloud deeper in the atmosphere to partially obscure the water vapour flux windows at $1.05\,\mathrm{\mu m}$ and $1.5\,\mathrm{\mu m}$.   While these partially cloudy models show reasonable agreement in the IR, it remains that none of the models presented here show good agreement with observations in the optical.   As discussed in \citet{lines_2019}, the agreement in the optical may be improved by increasing the width of the particle size distribution,  allowing for more small particles to scatter in the optical while maintaining the flat distribution in the IR.   In the context of {\sc EddySed}, this can be achieved by an increase in $\sigma$ which both widens the log-normal distribution and shifts the peak $r_\mathrm{g}$ to smaller values (see equation \ref{Eqn:rg}), although moving beyond a log-normal distribution to distributions better informed by microphysical modeling of clouds (e.g., \citealt{Powell_2019}) may be necessary. Unmodelled physics, such as submicron photochemical hazes, may also be necessary to improve the agreement with the transit spectrum in the UV and optical.  \citet{lavvas_2017} proposed, in their own attempts to address similar poor agreement in the optical between their model and the observed transit of HD189733b, that a reduction of the $\mathrm{H_2O}$ mixing ratio may be used to increase the effective radii  the UV and optical transit spectra relative to those in the IR.  We investigated this possibility in our own model via changes to the $\mathrm{H_2O}$ mixing ratio when generating post-processed transits and found that the effect was dependent on normalizing the spectra to the peak of water features and when normalizing instead to water flux windows (for example, normalizing the spectra to agree with observations at $1.3\, \mathrm{\mu m}$), the shift in the optical disappears.  Moreover, any change in the $\mathrm{H_2O}$ mixing ratio results in poor agreement in the IR where our models already agree with the prominent observed water features.

\begin{figure*}
	\includegraphics[]{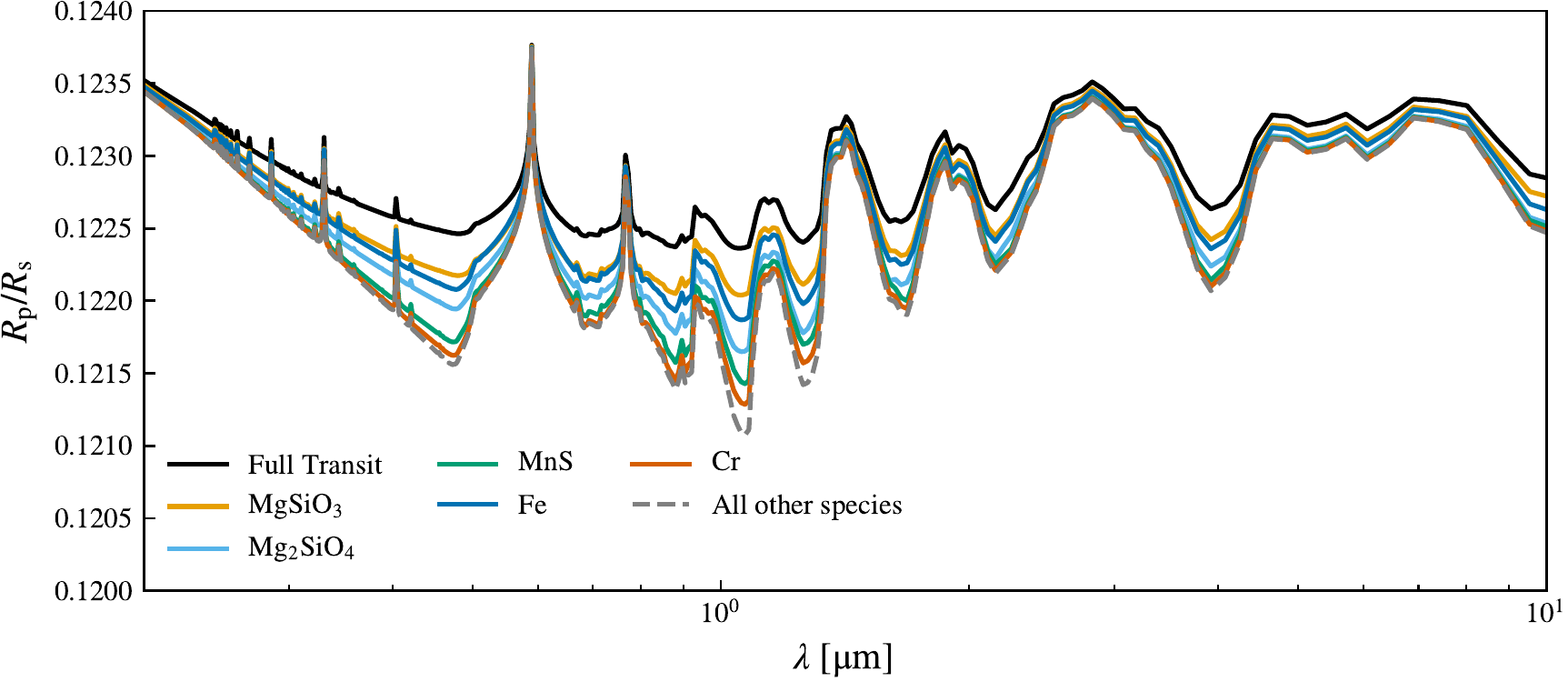}
    \caption{The post-processed transmission spectra for the P13 $f_\mathrm{sed}=2$ simulation. The black line shows the transmission spectrum including contributions from all cloud species, while the remaining lines show spectra taking into account clouds of a single species. The individual contributions of any species which contributes negligibly to the transmission spectrum are shown in grey. }
    \label{Fig:TransitSpecies}
\end{figure*}

The primary sources of cloud opacity in our transmission spectra are $\mathrm{MgSiO_3}$, $\mathrm{Mg_2SiO_4}$, and $\mathrm{Fe}$, with smaller contributions from $\mathrm{MnS}$ and $\mathrm{Cr}$, while all other cloud species contribute negligibly (see Fig. \ref{Fig:TransitSpecies}).  Although the condensation curve for $\mathrm{Mg_2SiO_4}$ used in EddySed does attempt to account for $\mathrm{MgSiO_3}$ and $\mathrm{Mg_2SiO_4}$ sharing gas phase precursors \citep{visscher_2010}, the relative contributions of these two cloud species is subject to an extra level of uncertainty.  

%A previous study using a microphysical model by \citet{2018MNRAS.481..194L} observed a prominent silicate feature between $8$ and $12\,\mathrm{\mu m}$ due to the cloud condensates, something not seen in the {\sc EddySed} modelling of \citet{lines_2019}.  The P13 models used here similarly do not show silicate features in the spectrum, with the likely cause being the use of a log-normal distribution in particle sizes in both {\sc EddySed} studies as opposed to using  a single representative particle size as in \citet{2018MNRAS.481..194L}.  The impact of distribution choice has been discussed previously in \citet{Powell_2019}.

\begin{figure*}
	\includegraphics[]{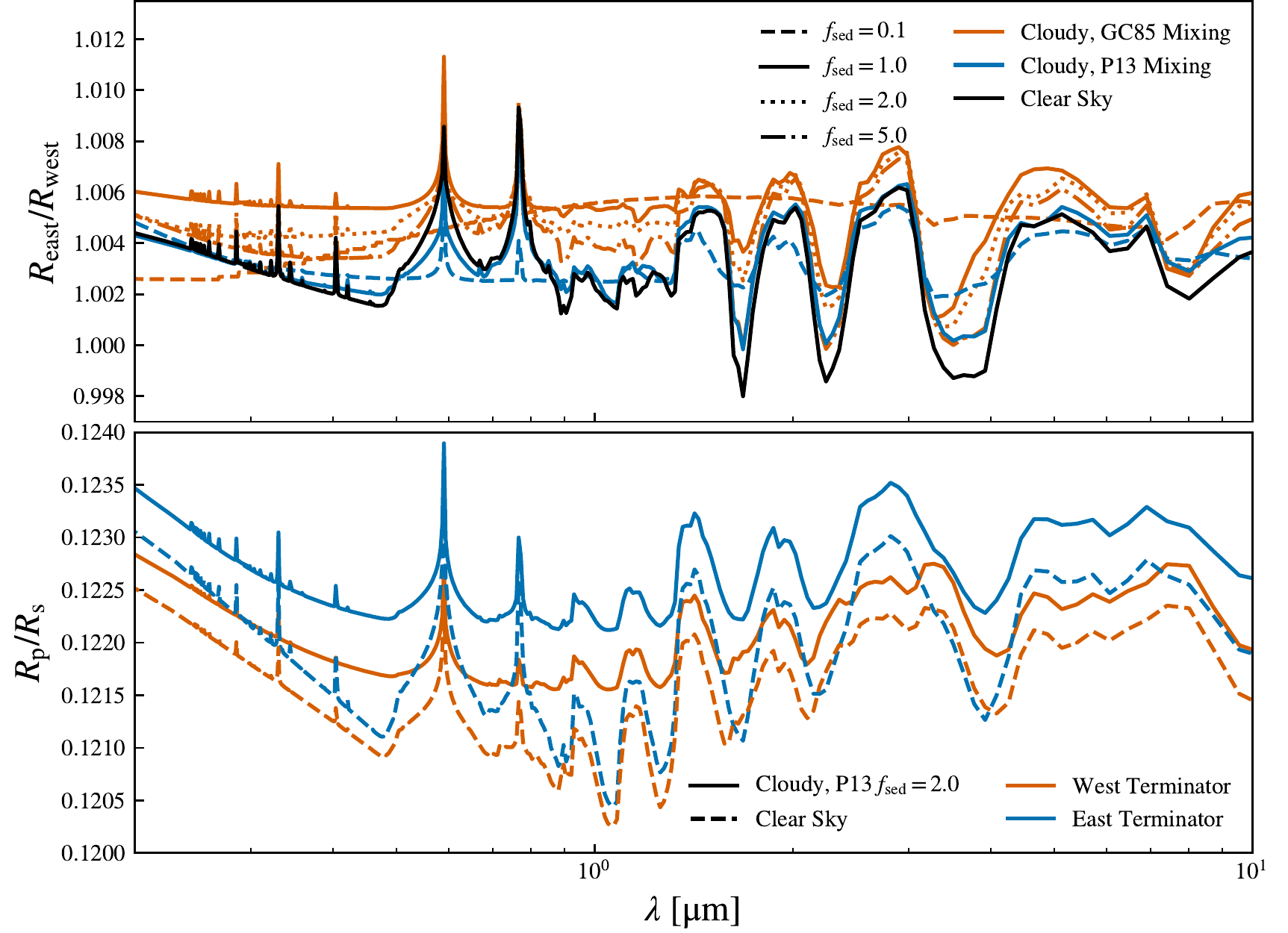}
    \caption{{\em Top:} The ratio of the inferred radii from the east and west limbs for all simulations with GC85 mixing (blue lines) and P13 mixing (red lines). The clear sky spectra is in black. For the cloudy cases, the linestyle indicates sedimentation factor.   {\em Bottom: }The east and west transmission spectra for the P13 $f_\mathrm{sed}=2.0$ and clear sky cases. }
    \label{Fig:TransitComp}
\end{figure*}

In order to further explore the asymmetry between the limbs of the transit, we separate each transit into its east and west components, including the poles, and  compute an effective radius.  Fig. \ref{Fig:TransitComp} (top panel) shows the ratio of these two radii for each model.  All models show a modest increase ($\lesssim 1\%$) in effective radius on the eastern limb.   A small exception to this occurs in the water flux windows between $1.4 \, \mathrm{\mu m}$ and $4\, \mathrm{\mu m}$ in the clear sky case where we see $R_\mathrm{east} \leq R_\mathrm{west}$.  The cause can be understood by looking at east and west transits individually. The bottom panel of Fig. \ref{Fig:TransitComp} shows the east and west transits for the clear sky case (dashed lines) as well as for the P13 $f_\mathrm{sed}=2.0$ case, chosen to be a representative model with a moderate amount of cloud.   In the clear sky case, it can be seen that the minima associated with the water vapour windows are offset between the east and west cases, and in some cases, the water features on the western limb are wider.   These two effects combine to create the regions of the spectra where the west limb is larger than the east limb.

In the optical, the spectral features seen in the transits (see Fig. \ref{Fig:Transit}) are also seen in $R_\mathrm{east}/R_\mathrm{west}$ with the greatest asymmetries being seen in the sodium and potassium lines in the optical and near-IR.  The P13 and GC85 models differ in that the P13 models, in the optical, show consistently greater asymmetry as $f_\mathrm{sed}$ decreases.

\subsubsection{Emission}
\label{sec:emission}

\begin{figure}
	\includegraphics[width=\columnwidth]{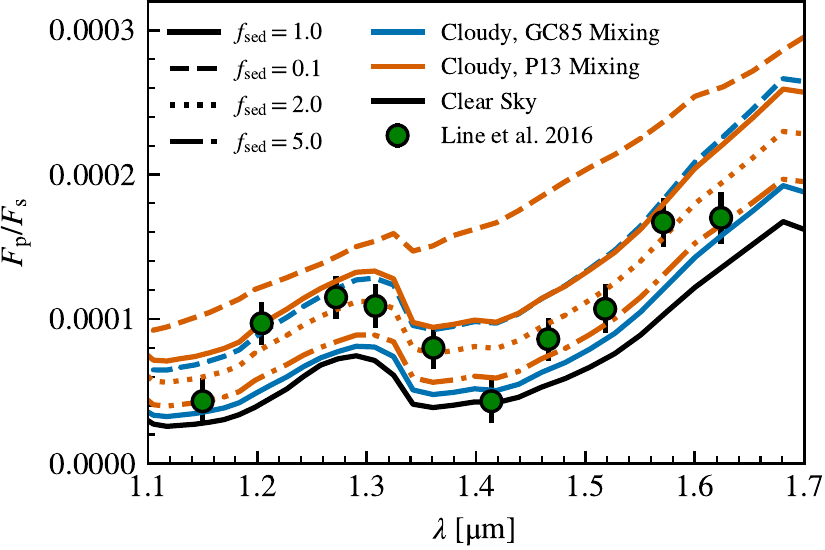}
    \caption{Dayside emission at WFC3 G141 1.1 - 1.7 $\mathrm{\mu m}$ for all cases.  Observational data from \citet{2016AJ....152..203L} are in green.}
    \label{Fig:Dayside-WFC3}
\end{figure}

\begin{figure}
	\includegraphics[width=\columnwidth]{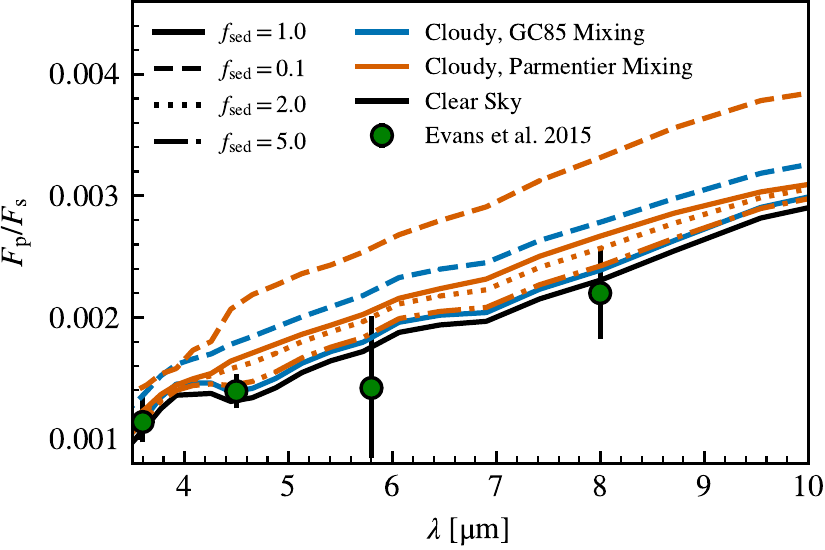}
    \caption{Dayside emission at 3.5 - 10 $\mathrm{\mu m}$ for all cases.  Observational data from \citet{evans_2015} are in green.}
    \label{Fig:Dayside2}
\end{figure}

Figs. \ref{Fig:Dayside-WFC3} and  \ref{Fig:Dayside2} show the dayside emission for the IR in the {\em Hubble} WFC3 G141 ($1.1$-$1.7 \,\mathrm{\mu m}$) and {\em Spitzer}/IRAC ($3.5$-$10\,\mathrm{\mu m}$) bands, respectively.  In each case, the flux shown is the sum of the reflected stellar component and the thermal planetary component.

We first examine the dayside emission in the near–IR WFC3 G141 ($1.1$-$1.7 \,\mathrm{\mu m}$) bandpass. Fig. \ref{Fig:Dayside-WFC3} shows the dayside emission for each run with the observations from \citet{2016AJ....152..203L} shown in green.  In general, the presence of cloud increases the dayside emission in this band, with the emission increasing with cloud vertical extent (i.e., lower $f_\mathrm{sed}$).   Of the GC85 simulations investigated by \citet{lines_2019}, the $f_\mathrm{sed} = 0.1$ case provides the best agreement with the data; however, with the inclusion of the P13 models, we now see better agreement with the P13 $f_\mathrm{sed} = 2.0$ model.  As was the case with the GC85 $f_\mathrm{sed} = 0.1$ model, the data point at $1.41\,\mathrm{\mu m}$ remains an outlier as it appears to be most consistent with the clear sky case.  While it may be tempting to point to this improved agreement with the observational data as due to the improved mixing treatment, it may simply be that our greater coverage of the $f_\mathrm{sed}$ parameter space for the P13 model led to an improved fit.   Since most of the data are bounded by the GC85 $f_\mathrm{sed} = 0.1$ and $1.0$ simulations, it is conceivable that an GC85 simulation with intermediate values of $f_\mathrm{sed}$ would yield an improved match to the data.  As our goal here is to investigate the P13 mixing treatment, a retrieval-type analysis optimizing parameters is beyond the scope of this particular work, and reserved for a future study.

Fig. \ref{Fig:Dayside2} shows the synthetic observations at 3.5 - 10 $\mathrm{\mu m}$ as well as the {\em Spitzer}/IRAC data of \citet{evans_2015} and \citet{zellem_2014}. The data are matched best by the clear sky case, although cloudy simulations with limited vertical extent (e.g., GC85 $f_\mathrm{sed} = 1.0$ and P13 $f_\mathrm{sed} = 5.0$) do not differ significantly from the clear sky spectrum.  This is consistent with the trend seen in the previous section where the data show the best agreement  with simulation with cloud scale heights less than $H$.

\subsubsection{Phase Curves}
\label{sec:phase}

\begin{figure}
	\includegraphics[]{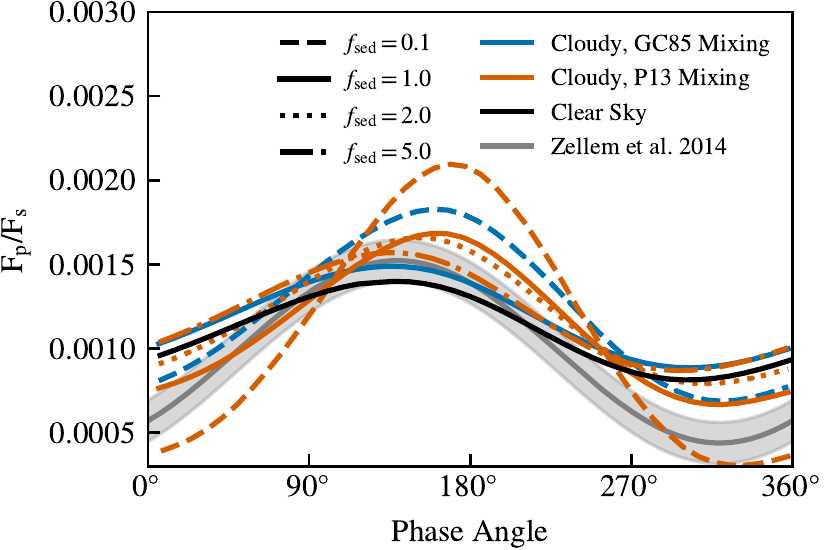}
    \caption{Phase curves at 4.5 $\mathrm{\mu m}$ for all cases.  A fit to the observational data from \citet{zellem_2014} is in grey with the shaded region indicating the $1\sigma$ error.}
    \label{Fig:phase45micron}
\end{figure}

The 4.5 micron phase curves for all the models are shown in Fig. \ref{Fig:phase45micron} with the observed phase curve from \cite{zellem_2014} in  grey.  The clear sky model exhibits a larger nightside flux compared to observations, and in general underpredicts the contrast between the day and night sides, an effect for cloudless atmospheres that is well documented in the literature \citep{showman_2009,parmentier_2016,parmentier_2020}.  Both mixing treatments increase the contrast in fluxes between the dayside and the nightside; however, with the exception of the P13 $f_\mathrm{sed} = 0.1$ case, they all continue to overestimate the nightside flux.  The dayside flux, on the other hand, shows best agreement with the models with limited cloud (GC85 $f_\mathrm{sed}=1.0$ and P13 $f_\mathrm{sed}=5.0$) as discussed in the Section \ref{sec:emission}, so it unlikely that the  observed $4.5\,\mathrm{\mu m}$ phase curve can be explained through variation of $f_\mathrm{sed}$ alone.  Parameters not varied within this study, such as the width of the log-normal distribution or the specific condensate species included, may provide avenues to resolve the current discrepancies.

%\begin{figure}
%	\includegraphics[width=\columnwidth]{offset.pdf}
%    \caption{The dayside hot spot offset relative to the substellar point for the HDI runs.  The dashed line indicates the  The hot spot peak is taken to be at the location of the maximum upward thermal flux on the outer boundary of the simulation.  }
%    \label{Fig:Offset}
%\end{figure}

%\begin{figure*}
%	\includegraphics[width=\textwidth]{zonalvel_panel.pdf}
%    \caption{Zonal Wind at 0.01 bar (left column), 0.1 bar (middle column) and 1 bar (right column) for the HDI runs.}
%    \label{Fig:ZVelPanel}
%\end{figure*}

%\begin{figure*}
%	\includegraphics[width=\textwidth]{wvel_panel.pdf}
%    \caption{Vertical velocity at 0.01 bar (left column), 0.1 bar (middle column) and 1 bar (right column) for the HDI runs.}
%    \label{Fig:WVelPanel}
%\end{figure*}

\section{Conclusions}
\label{sec:conclusions}

In this work we have investigated the impact of a more physically accurate mixing treatment on the {\sc EddySed} cloud model when applied to hot Jupiters.  The default convective mixing of \citet{1985rapm.book..121G} (referred to here at GC85) is replaced with a mixing treatment based on \citet{2013A&A...558A..91P} (P13) which explicitly models mixing in the atmosphere of HD209458b through tracers within a GCM.  We performed simulations using the new P13 mixing treatment for four values of the sedimentation factor $f_\mathrm{sed}$ covering effective cloud scale heights from $0.2H$ to $10H$.  

Within the {\sc EddySed} formalism, the sedimentation factor $f_\mathrm{sed}$ and the mixing length $L_\mathrm{mix}$ are the primary drivers of the cloud distribution as they form the cloud scale $L_\mathrm{cloud} = f_\mathrm{sed}^{-1}L_\mathrm{mix}$ , with the eddy mixing rate $K_{zz}$ influencing the distribution only indirectly through its role in determining the particle sizes and optical properties. As a result, we find that the choice of sedimentation factor -- or equivalently, the choice of cloud scale --  plays a larger role in the atmospheric evolution than the switch in choice of $K_{zz}$ given the relative unconstrained nature of $f_\text{sed}$.  We do observe a decrease in particle size due to the switch to the P13 mixing treatment; however, due to the relatively weak scaling of $r_\mathrm{g}$ with $K_{zz}$, the change was insufficient to result in significant differences in observables.  While not investigated here,  the width of the log-normal particle size distribution -- a fixed parameter in {\sc EddySed} -- may impact results to a greater degree due to the sensitivity of the peak radius $r_\text{g}$ to the width $\sigma$.    We also observe that differences in chosen mixing scale $L_\mathrm{mix}$ can roughly be accounted for by a commensurate scaling of the sedimentation factor.   The results of our GC85 $f_\mathrm{sed} = 1.0$ and P13 $f_\mathrm{sed} = 5.0$ simulations, which differ by a factor of 5 in sedimentation factor as well as in average mixing scale, show very similar atmospheric structures as well as transits and emission spectra due to both models having average cloud scales $\left<L_\mathrm{cloud}/H\right>_V = 0.2$. This rough equivalence may not hold for simulations with larger differences in $K_{zz}$ or for cases with differing distribution widths.

In the P13 cases, we see qualitatively the same distribution of condensates as in the GC85 models, with clouds forming at higher pressures within the atmosphere near the poles, and sedimentation reducing the vapour concentrations above the cloud layers.  This should be noted as it contrasts with purely temperature-dependent models which use a constant vapour concentration throughout the atmosphere which may result in larger vertical extents in the resulting clouds.   We see the greatest amount of condensate at low pressure near the poles in the P13 $f_\mathrm{sed}=0.1$  model which has a corresponding cloud scale of $L_\mathrm{cloud}=10H$.   Near the equator, we find that the higher temperatures, especially on the dayside, push the cloud base to lower pressures, with the cloud forming an ``arch'' over the hottest parts of the equatorial jet.   These morphological differences are due to the approximate modelling of sedimentation and mixing in {\sc EddySed} are taken to be improvements over more simplified models.

Observationally, we find that our parameter study using the P13 mixing treatment better agrees with the observed transits and dayside emission in the WFC3 G141 bandpass, specifically the simulation with $f_\mathrm{sed} = 2.0$; however, the parameter study covers intermediate cloud scales not seen in the GC85 parameter study done by \citet{lines_2019}, and it may be possible to find good agreement with the data using intermediate values of $f_\mathrm{sed}$ between $0.1$ and $1.0$. Neither the GC85 nor the P13 simulations were able to reproduce the $4.5\,\mathrm{\mu m}$ phase curve with most models overestimating the nightside flux; however, the cloud cases showed better agreement with observations than the clear sky case.  Similarly, the shallow cloud cases (P13 $f_\mathrm{sed} = 2.0$, specifically) improved the agreement with the IR transit over clear sky case.   It remains that neither the cloud nor the clear sky cases show good agreement in the optical and UV.   

Our simulated transits did show evidence for a small increase ($\lesssim 1\% $) in effective radius in the eastern limb compared to the western limb, with simulations with the largest cloud scales showing the greatest asymmetry in the optical across the band.  $\text{Na}$ and $\text{K}$ lines in the optical and near-IR are more prominent in the eastern limb resulting in a local increase in relative radius around these lines.  In the IR, we find $\mathrm{H_2O}$ and $\text{CO}$ features cause large variations in the relative radius although the eastern radius remaining larger.  

The agreement with observations may be improved by examining parameters within the {\sc EddySed} model not varied in this study.  We adopted a fixed width for the log-normal particle size distribution, and as was discussed earlier and in \citet{lines_2019}, a wider distribution would increase the number of small particles scattering in the optical which may improve the fit to the observed transmission spectrum.    We have also opted not to investigate which species contribute to cloud formation, instead allowing all species for which we have condensation curves to potentially condense into clouds.  Excluding or limiting species with large energy barriers such as $\mathrm{MnS}$ or $\mathrm{Fe}$ may improve our agreement with observations.   It may be the case, however, that  physics beyond what is modelled in {\sc EddySed} is necessary.  

While it remains that the simulations presented here are unable to explain all observations, they do represent an improvement over more simplified models with a limited increase in computational overhead, especially compared to true microphysical models.  In the future, we hope to study the impact of the particle size distribution on the results, something that has been overlooked in this and previous {\sc EddySed} studies.   In the longer term, we hope to adapt and expand the model to account for the local velocity field and address the shortcomings of using the global mixing treatment.

\section*{Acknowledgements}

This research made use of the ISCA High Performance Computing Service at the University of Exeter. This work was performed using the DiRAC Data Intensive service at Leicester, operated by the University of Leicester IT Services, which forms part of the STFC DiRAC HPC Facility (www.dirac.ac.uk). The equipment was funded by BEIS capital funding via STFC capital grants ST/K000373/1 and ST/R002363/1 and STFC DiRAC Operations grant ST/R001014/1. DiRAC is part of the National e-Infrastructure. This work was partly supported by a Science and Technology Facilities Council Consolidated Grant (ST/R000395/1), and by the Leverhulme Trust through a research project grant (RPG-2020-82). This research was also made possible by funding through the UKRI Future Leaders Scheme: MR/T040866/1. Material produced using Met Office Software. We acknowledge use of the Monsoon system, a collaborative facility supplied under the Joint Weather and Climate Research Programme, a strategic partnership between the Met Office and the Natural Environment Research Council. This work benefited from the 2018 Exoplanet Summer Program in the Other Worlds Laboratory (OWL) at the University of California, Santa Cruz, a program funded by the Heising-Simons Foundation. This work benefited from crucial discussions with Mark Marley, Andy Ackerman and Natasha Batalha, alongside provision of software ({\sc EddySed}) by Tiffany Kataria.   Additionally, the analysis of simulation data and creation of plots were done using the invaluable Python packages {\sc Numpy} \citep{numpy} and {\sc Matplotlib} \citep{matplotlib}.

%Carbon Footprint
We would also like to note the energy intensive nature of supercomputing and the greenhouse gas emissions associated with it.  We estimate the final production runs needed for this paper resulted in roughly 0.5 tonnes of $\mathrm{CO_2}$ emitted into the atmosphere, which does not include emissions associated with testing and development.

\section*{Data Availability}

The research data supporting this publication are openly available at https://doi.org/10.7910/DVN/NLT6MS. 

%%%%%%%%%%%%%%%%%%%%%%%%%%%%%%%%%%%%%%%%%%%%%%%%%%



\bibliographystyle{mnras}
\bibliography{bibliography} 

\begin{thebibliography}{}
\makeatletter
\relax
\def\mn@urlcharsother{\let\do\@makeother \do\$\do\&\do\#\do\^\do\_\do\%\do\~}
\def\mn@doi{\begingroup\mn@urlcharsother \@ifnextchar [ {\mn@doi@}
  {\mn@doi@[]}}
\def\mn@doi@[#1]#2{\def\@tempa{#1}\ifx\@tempa\@empty \href
  {http://dx.doi.org/#2} {doi:#2}\else \href {http://dx.doi.org/#2} {#1}\fi
  \endgroup}
\def\mn@eprint#1#2{\mn@eprint@#1:#2::\@nil}
\def\mn@eprint@arXiv#1{\href {http://arxiv.org/abs/#1} {{\tt arXiv:#1}}}
\def\mn@eprint@dblp#1{\href {http://dblp.uni-trier.de/rec/bibtex/#1.xml}
  {dblp:#1}}
\def\mn@eprint@#1:#2:#3:#4\@nil{\def\@tempa {#1}\def\@tempb {#2}\def\@tempc
  {#3}\ifx \@tempc \@empty \let \@tempc \@tempb \let \@tempb \@tempa \fi \ifx
  \@tempb \@empty \def\@tempb {arXiv}\fi \@ifundefined
  {mn@eprint@\@tempb}{\@tempb:\@tempc}{\expandafter \expandafter \csname
  mn@eprint@\@tempb\endcsname \expandafter{\@tempc}}}

\bibitem[\protect\citeauthoryear{{Ackerman} \& {Marley}}{{Ackerman} \&
  {Marley}}{2001}]{2001ApJ...556..872A}
{Ackerman} A.~S.,  {Marley} M.~S.,  2001, \mn@doi [\apj] {10.1086/321540},
  \href {https://ui.adsabs.harvard.edu/abs/2001ApJ...556..872A} {556, 872}

\bibitem[\protect\citeauthoryear{{Amundsen}, {Baraffe}, {Tremblin}, {Manners},
  {Hayek}, {Mayne}  \& {Acreman}}{{Amundsen}
  et~al.}{2014}]{2014A&A...564A..59A}
{Amundsen} D.~S.,  {Baraffe} I.,  {Tremblin} P.,  {Manners} J.,  {Hayek} W.,
  {Mayne} N.~J.,   {Acreman} D.~M.,  2014, \mn@doi [\aap]
  {10.1051/0004-6361/201323169}, \href
  {https://ui.adsabs.harvard.edu/abs/2014A&A...564A..59A} {564, A59}

\bibitem[\protect\citeauthoryear{{Amundsen} et~al.,}{{Amundsen}
  et~al.}{2016}]{2016A&A...595A..36A}
{Amundsen} D.~S.,  et~al., 2016, \mn@doi [\aap] {10.1051/0004-6361/201629183},
  \href {https://ui.adsabs.harvard.edu/abs/2016A&A...595A..36A} {595, A36}

\bibitem[\protect\citeauthoryear{{Amundsen}, {Tremblin}, {Manners}, {Baraffe}
  \& {Mayne}}{{Amundsen} et~al.}{2017}]{amundsen_2017}
{Amundsen} D.~S.,  {Tremblin} P.,  {Manners} J.,  {Baraffe} I.,   {Mayne}
  N.~J.,  2017, \mn@doi [\aap] {10.1051/0004-6361/201629322}, \href
  {https://ui.adsabs.harvard.edu/abs/2017A&A...598A..97A} {598, A97}

\bibitem[\protect\citeauthoryear{{Armstrong}, {de Mooij}, {Barstow}, {Osborn},
  {Blake}  \& {Saniee}}{{Armstrong} et~al.}{2016}]{armstrong_2016}
{Armstrong} D.~J.,  {de Mooij} E.,  {Barstow} J.,  {Osborn} H.~P.,  {Blake} J.,
    {Saniee} N.~F.,  2016, \mn@doi [Nature Astronomy]
  {10.1038/s41550-016-0004}, \href
  {http://adsabs.harvard.edu/abs/2016NatAs...1E...4A} {1, 0004}

\bibitem[\protect\citeauthoryear{{Baeyens}, {Decin}, {Carone}, {Venot},
  {Ag{\'u}ndez}  \& {Molli{\'e}re}}{{Baeyens} et~al.}{2021}]{baeyens_2021}
{Baeyens} R.,  {Decin} L.,  {Carone} L.,  {Venot} O.,  {Ag{\'u}ndez} M.,
  {Molli{\'e}re} P.,  2021, \mn@doi [\mnras] {10.1093/mnras/stab1310}, \href
  {https://ui.adsabs.harvard.edu/abs/2021MNRAS.tmp.1277B} {}

\bibitem[\protect\citeauthoryear{{Barstow}, {Aigrain}, {Irwin}  \&
  {Sing}}{{Barstow} et~al.}{2017}]{Barstow_2017}
{Barstow} J.~K.,  {Aigrain} S.,  {Irwin} P.~G.~J.,   {Sing} D.~K.,  2017,
  \mn@doi [\apj] {10.3847/1538-4357/834/1/50}, \href
  {https://ui.adsabs.harvard.edu/abs/2017ApJ...834...50B} {834, 50}

\bibitem[\protect\citeauthoryear{{Brown}}{{Brown}}{2001}]{brown_2001}
{Brown} T.~M.,  2001, \mn@doi [\apj] {10.1086/320950}, \href
  {https://ui.adsabs.harvard.edu/abs/2001ApJ...553.1006B} {553, 1006}

\bibitem[\protect\citeauthoryear{{Bruno} et~al.,}{{Bruno}
  et~al.}{2018}]{bruno_2018}
{Bruno} G.,  et~al., 2018, \mn@doi [\aj] {10.3847/1538-3881/aaa0c7}, \href
  {https://ui.adsabs.harvard.edu/abs/2018AJ....155...55B} {155, 55}

\bibitem[\protect\citeauthoryear{{Burrows} \& {Sharp}}{{Burrows} \&
  {Sharp}}{1999}]{1999ApJ...512..843B}
{Burrows} A.,  {Sharp} C.~M.,  1999, \mn@doi [\apj] {10.1086/306811}, \href
  {https://ui.adsabs.harvard.edu/abs/1999ApJ...512..843B} {512, 843}

\bibitem[\protect\citeauthoryear{{Chachan} et~al.,}{{Chachan}
  et~al.}{2019}]{2019AJ....158..244C}
{Chachan} Y.,  et~al., 2019, \mn@doi [\aj] {10.3847/1538-3881/ab4e9a}, \href
  {https://ui.adsabs.harvard.edu/abs/2019AJ....158..244C} {158, 244}

\bibitem[\protect\citeauthoryear{{Charnay}, {B{\'e}zard}, {Baudino},
  {Bonnefoy}, {Boccaletti}  \& {Galicher}}{{Charnay}
  et~al.}{2018}]{charnay_2018}
{Charnay} B.,  {B{\'e}zard} B.,  {Baudino} J.~L.,  {Bonnefoy} M.,  {Boccaletti}
  A.,   {Galicher} R.,  2018, \mn@doi [\apj] {10.3847/1538-4357/aaac7d}, \href
  {https://ui.adsabs.harvard.edu/abs/2018ApJ...854..172C} {854, 172}

\bibitem[\protect\citeauthoryear{{Dang} et~al.,}{{Dang}
  et~al.}{2018}]{dang_2018}
{Dang} L.,  et~al., 2018, \mn@doi [Nature Astronomy]
  {10.1038/s41550-017-0351-6}, \href
  {https://ui.adsabs.harvard.edu/abs/2018NatAs...2..220D} {2, 220}

\bibitem[\protect\citeauthoryear{{Deming} et~al.,}{{Deming}
  et~al.}{2013}]{2013ApJ...774...95D}
{Deming} D.,  et~al., 2013, \mn@doi [\apj] {10.1088/0004-637X/774/2/95}, \href
  {https://ui.adsabs.harvard.edu/abs/2013ApJ...774...95D} {774, 95}

\bibitem[\protect\citeauthoryear{{Demory} et~al.,}{{Demory}
  et~al.}{2011}]{demory_2011}
{Demory} B.-O.,  et~al., 2011, \mn@doi [\apjl] {10.1088/2041-8205/735/1/L12},
  \href {http://adsabs.harvard.edu/abs/2011ApJ...735L..12D} {735, L12}

\bibitem[\protect\citeauthoryear{{Dobbs-Dixon} \& {Agol}}{{Dobbs-Dixon} \&
  {Agol}}{2013}]{dobbs_dixon_2013}
{Dobbs-Dixon} I.,  {Agol} E.,  2013, \mn@doi [\mnras] {10.1093/mnras/stt1509},
  \href {https://ui.adsabs.harvard.edu/abs/2013MNRAS.435.3159D} {435, 3159}

\bibitem[\protect\citeauthoryear{{Drummond}, {Tremblin}, {Baraffe}, {Amundsen},
  {Mayne}, {Venot}  \& {Goyal}}{{Drummond} et~al.}{2016}]{drummond_2016}
{Drummond} B.,  {Tremblin} P.,  {Baraffe} I.,  {Amundsen} D.~S.,  {Mayne}
  N.~J.,  {Venot} O.,   {Goyal} J.,  2016, \mn@doi [\aap]
  {10.1051/0004-6361/201628799}, \href
  {https://ui.adsabs.harvard.edu/abs/2016A&A...594A..69D} {594, A69}

\bibitem[\protect\citeauthoryear{{Drummond}, {Mayne}, {Baraffe}, {Tremblin},
  {Manners}, {Amundsen}, {Goyal}  \& {Acreman}}{{Drummond}
  et~al.}{2018a}]{drummond_2018b}
{Drummond} B.,  {Mayne} N.~J.,  {Baraffe} I.,  {Tremblin} P.,  {Manners} J.,
  {Amundsen} D.~S.,  {Goyal} J.,   {Acreman} D.,  2018a, \mn@doi [\aap]
  {10.1051/0004-6361/201732010}, \href
  {https://ui.adsabs.harvard.edu/abs/2018A&A...612A.105D} {612, A105}

\bibitem[\protect\citeauthoryear{{Drummond} et~al.,}{{Drummond}
  et~al.}{2018b}]{drummond_2018a}
{Drummond} B.,  et~al., 2018b, \mn@doi [\apjl] {10.3847/2041-8213/aab209},
  \href {https://ui.adsabs.harvard.edu/abs/2018ApJ...855L..31D} {855, L31}

\bibitem[\protect\citeauthoryear{{Drummond}, {Mayne}, {Manners}, {Baraffe},
  {Goyal}, {Tremblin}, {Sing}  \& {Kohary}}{{Drummond}
  et~al.}{2018c}]{drummond_2018c}
{Drummond} B.,  {Mayne} N.~J.,  {Manners} J.,  {Baraffe} I.,  {Goyal} J.,
  {Tremblin} P.,  {Sing} D.~K.,   {Kohary} K.,  2018c, \mn@doi [\apj]
  {10.3847/1538-4357/aaeb28}, \href
  {https://ui.adsabs.harvard.edu/abs/2018ApJ...869...28D} {869, 28}

\bibitem[\protect\citeauthoryear{{Drummond} et~al.,}{{Drummond}
  et~al.}{2020}]{drummond_2020}
{Drummond} B.,  et~al., 2020, \mn@doi [\aap] {10.1051/0004-6361/201937153},
  \href {https://ui.adsabs.harvard.edu/abs/2020A&A...636A..68D} {636, A68}

\bibitem[\protect\citeauthoryear{{Edwards} \& {Slingo}}{{Edwards} \&
  {Slingo}}{1996}]{1996QJRMS.122..689E}
{Edwards} J.~M.,  {Slingo} A.,  1996, \mn@doi [Quarterly Journal of the Royal
  Meteorological Society] {10.1002/qj.49712253107}, \href
  {https://ui.adsabs.harvard.edu/abs/1996QJRMS.122..689E} {122, 689}

\bibitem[\protect\citeauthoryear{{Evans}, {Aigrain}, {Gibson}, {Barstow},
  {Amundsen}, {Tremblin}  \& {Mourier}}{{Evans} et~al.}{2015}]{evans_2015}
{Evans} T.~M.,  {Aigrain} S.,  {Gibson} N.,  {Barstow} J.~K.,  {Amundsen}
  D.~S.,  {Tremblin} P.,   {Mourier} P.,  2015, \mn@doi [\mnras]
  {10.1093/mnras/stv910}, \href
  {https://ui.adsabs.harvard.edu/abs/2015MNRAS.451..680E} {451, 680}

\bibitem[\protect\citeauthoryear{{Feng}, {Line}  \& {Fortney}}{{Feng}
  et~al.}{2020}]{feng_2020}
{Feng} Y.~K.,  {Line} M.~R.,   {Fortney} J.~J.,  2020, \mn@doi [\aj]
  {10.3847/1538-3881/aba8f9}, \href
  {https://ui.adsabs.harvard.edu/abs/2020AJ....160..137F} {160, 137}

\bibitem[\protect\citeauthoryear{{Fisher} \& {Heng}}{{Fisher} \&
  {Heng}}{2018}]{Fisher_2018}
{Fisher} C.,  {Heng} K.,  2018, \mn@doi [\mnras] {10.1093/mnras/sty2550}, \href
  {https://ui.adsabs.harvard.edu/abs/2018MNRAS.481.4698F} {481, 4698}

\bibitem[\protect\citeauthoryear{{Fu}, {Deming}, {Knutson}, {Madhusudhan},
  {Mandell}  \& {Fraine}}{{Fu} et~al.}{2017}]{fu_2017}
{Fu} G.,  {Deming} D.,  {Knutson} H.,  {Madhusudhan} N.,  {Mandell} A.,
  {Fraine} J.,  2017, \mn@doi [\apjl] {10.3847/2041-8213/aa8e40}, \href
  {https://ui.adsabs.harvard.edu/abs/2017ApJ...847L..22F} {847, L22}

\bibitem[\protect\citeauthoryear{{Gao}, {Marley}  \& {Ackerman}}{{Gao}
  et~al.}{2018}]{gao_2018}
{Gao} P.,  {Marley} M.~S.,   {Ackerman} A.~S.,  2018, \mn@doi [\apj]
  {10.3847/1538-4357/aab0a1}, \href
  {https://ui.adsabs.harvard.edu/abs/2018ApJ...855...86G} {855, 86}

\bibitem[\protect\citeauthoryear{{Gierasch} \& {Conrath}}{{Gierasch} \&
  {Conrath}}{1985}]{1985rapm.book..121G}
{Gierasch} P.~J.,  {Conrath} B.~J.,  1985, {Energy conversion processes in the
  outer planets.}.
pp 121--146

\bibitem[\protect\citeauthoryear{{Goyal} et~al.,}{{Goyal}
  et~al.}{2018}]{goyal_2018}
{Goyal} J.~M.,  et~al., 2018, \mn@doi [\mnras] {10.1093/mnras/stx3015}, \href
  {https://ui.adsabs.harvard.edu/abs/2018MNRAS.474.5158G} {474, 5158}

\bibitem[\protect\citeauthoryear{{Harada}, {Kempton}, {Rauscher}, {Roman}  \&
  {Brinjikji}}{{Harada} et~al.}{2019}]{harada_2019}
{Harada} C.~K.,  {Kempton} E. M.~R.,  {Rauscher} E.,  {Roman} M.,   {Brinjikji}
  M.,  2019, arXiv e-prints, \href
  {https://ui.adsabs.harvard.edu/abs/2019arXiv191202268H} {p. arXiv:1912.02268}

\bibitem[\protect\citeauthoryear{Harris et~al.,}{Harris et~al.}{2020}]{numpy}
Harris C.~R.,  et~al., 2020, \mn@doi [Nature] {10.1038/s41586-020-2649-2}, 585,
  357

\bibitem[\protect\citeauthoryear{{Helling} \& {Fomins}}{{Helling} \&
  {Fomins}}{2013}]{2013RSPTA.37110581H}
{Helling} C.,  {Fomins} A.,  2013, \mn@doi [Philosophical Transactions of the
  Royal Society of London Series A] {10.1098/rsta.2011.0581}, \href
  {https://ui.adsabs.harvard.edu/abs/2013RSPTA.37110581H} {371, 20110581}

\bibitem[\protect\citeauthoryear{{Helling} et~al.,}{{Helling}
  et~al.}{2016}]{helling_2016}
{Helling} C.,  et~al., 2016, \mn@doi [\mnras] {10.1093/mnras/stw662}, \href
  {https://ui.adsabs.harvard.edu/abs/2016MNRAS.460..855H} {460, 855}

\bibitem[\protect\citeauthoryear{{Helling}, {Gourbin}, {Woitke}  \&
  {Parmentier}}{{Helling} et~al.}{2019}]{Helling_2019}
{Helling} C.,  {Gourbin} P.,  {Woitke} P.,   {Parmentier} V.,  2019, \mn@doi
  [\aap] {10.1051/0004-6361/201834085}, \href
  {https://ui.adsabs.harvard.edu/abs/2019A&A...626A.133H} {626, A133}

\bibitem[\protect\citeauthoryear{{Heng}}{{Heng}}{2016}]{heng_2016}
{Heng} K.,  2016, \mn@doi [\apjl] {10.3847/2041-8205/826/1/L16}, \href
  {https://ui.adsabs.harvard.edu/abs/2016ApJ...826L..16H} {826, L16}

\bibitem[\protect\citeauthoryear{{Hindle}, {Bushby}  \& {Rogers}}{{Hindle}
  et~al.}{2019}]{hindle_2019}
{Hindle} A.~W.,  {Bushby} P.~J.,   {Rogers} T.~M.,  2019, \mn@doi [\apjl]
  {10.3847/2041-8213/ab05dd}, \href
  {https://ui.adsabs.harvard.edu/abs/2019ApJ...872L..27H} {872, L27}

\bibitem[\protect\citeauthoryear{{Hubbard}, {Fortney}, {Lunine}, {Burrows},
  {Sudarsky}  \& {Pinto}}{{Hubbard} et~al.}{2001}]{hubbard_2001}
{Hubbard} W.~B.,  {Fortney} J.~J.,  {Lunine} J.~I.,  {Burrows} A.,  {Sudarsky}
  D.,   {Pinto} P.,  2001, \mn@doi [\apj] {10.1086/322490}, \href
  {https://ui.adsabs.harvard.edu/abs/2001ApJ...560..413H} {560, 413}

\bibitem[\protect\citeauthoryear{Hunter}{Hunter}{2007}]{matplotlib}
Hunter J.~D.,  2007, \mn@doi [Computing in Science {\&} Engineering]
  {10.1109/MCSE.2007.55}, 9, 90

\bibitem[\protect\citeauthoryear{{Irwin}, {Parmentier}, {Taylor}, {Barstow},
  {Aigrain}, {Lee}  \& {Garland }}{{Irwin} et~al.}{2020}]{Irwin_2020}
{Irwin} P. G.~J.,  {Parmentier} V.,  {Taylor} J.,  {Barstow} J.,  {Aigrain} S.,
   {Lee} G. K.~H.,   {Garland } R.,  2020, \mn@doi [\mnras]
  {10.1093/mnras/staa238}, \href
  {https://ui.adsabs.harvard.edu/abs/2020MNRAS.493..106I} {493, 106}

\bibitem[\protect\citeauthoryear{{Iyer}, {Swain}, {Zellem}, {Line}, {Roudier},
  {Rocha}  \& {Livingston}}{{Iyer} et~al.}{2016}]{2016ApJ...823..109I}
{Iyer} A.~R.,  {Swain} M.~R.,  {Zellem} R.~T.,  {Line} M.~R.,  {Roudier} G.,
  {Rocha} G.,   {Livingston} J.~H.,  2016, \mn@doi [\apj]
  {10.3847/0004-637X/823/2/109}, \href
  {https://ui.adsabs.harvard.edu/abs/2016ApJ...823..109I} {823, 109}

\bibitem[\protect\citeauthoryear{{Juncher}, {J{\o}rgensen}  \&
  {Helling}}{{Juncher} et~al.}{2017}]{juncher_2017}
{Juncher} D.,  {J{\o}rgensen} U.~G.,   {Helling} C.,  2017, \mn@doi [\aap]
  {10.1051/0004-6361/201629977}, \href
  {https://ui.adsabs.harvard.edu/abs/2017A&A...608A..70J} {608, A70}

\bibitem[\protect\citeauthoryear{{Komacek}, {Showman}  \&
  {Parmentier}}{{Komacek} et~al.}{2019}]{komacek_2019}
{Komacek} T.~D.,  {Showman} A.~P.,   {Parmentier} V.,  2019, \mn@doi [\apj]
  {10.3847/1538-4357/ab338b}, \href
  {https://ui.adsabs.harvard.edu/abs/2019ApJ...881..152K} {881, 152}

\bibitem[\protect\citeauthoryear{{Kozasa}, {Hasegawa}  \& {Nomoto}}{{Kozasa}
  et~al.}{1989}]{kozasa_1989}
{Kozasa} T.,  {Hasegawa} H.,   {Nomoto} K.,  1989, \mn@doi [\apj]
  {10.1086/167801}, \href
  {https://ui.adsabs.harvard.edu/abs/1989ApJ...344..325K} {344, 325}

\bibitem[\protect\citeauthoryear{{Kreidberg} et~al.,}{{Kreidberg}
  et~al.}{2014}]{kreidberg_2014}
{Kreidberg} L.,  et~al., 2014, \mn@doi [\nat] {10.1038/nature12888}, \href
  {http://adsabs.harvard.edu/abs/2014Natur.505...69K} {505, 69}

\bibitem[\protect\citeauthoryear{{Kreidberg}, {Line}, {Thorngren}, {Morley}  \&
  {Stevenson}}{{Kreidberg} et~al.}{2018}]{kreidberg_2018}
{Kreidberg} L.,  {Line} M.~R.,  {Thorngren} D.,  {Morley} C.~V.,   {Stevenson}
  K.~B.,  2018, \mn@doi [\apjl] {10.3847/2041-8213/aabfce}, \href
  {https://ui.adsabs.harvard.edu/abs/2018ApJ...858L...6K} {858, L6}

\bibitem[\protect\citeauthoryear{{Lavvas} \& {Koskinen}}{{Lavvas} \&
  {Koskinen}}{2017}]{lavvas_2017}
{Lavvas} P.,  {Koskinen} T.,  2017, \mn@doi [\apj] {10.3847/1538-4357/aa88ce},
  \href {https://ui.adsabs.harvard.edu/abs/2017ApJ...847...32L} {847, 32}

\bibitem[\protect\citeauthoryear{{Lecavelier Des Etangs}, {Pont},
  {Vidal-Madjar}  \& {Sing}}{{Lecavelier Des Etangs}
  et~al.}{2008}]{2008A&A...481L..83L}
{Lecavelier Des Etangs} A.,  {Pont} F.,  {Vidal-Madjar} A.,   {Sing} D.,  2008,
  \mn@doi [\aap] {10.1051/0004-6361:200809388}, \href
  {https://ui.adsabs.harvard.edu/abs/2008A&A...481L..83L} {481, L83}

\bibitem[\protect\citeauthoryear{{Lee}, {Dobbs-Dixon}, {Helling}, {Bognar}  \&
  {Woitke}}{{Lee} et~al.}{2016}]{2016A&A...594A..48L}
{Lee} G.,  {Dobbs-Dixon} I.,  {Helling} C.,  {Bognar} K.,   {Woitke} P.,  2016,
  \mn@doi [\aap] {10.1051/0004-6361/201628606}, \href
  {https://ui.adsabs.harvard.edu/abs/2016A&A...594A..48L} {594, A48}

\bibitem[\protect\citeauthoryear{{Lindzen}}{{Lindzen}}{1981}]{1981JGR....86.9707L}
{Lindzen} R.~S.,  1981, \mn@doi [\jgr] {10.1029/JC086iC10p09707}, \href
  {https://ui.adsabs.harvard.edu/abs/1981JGR....86.9707L} {86, 9707}

\bibitem[\protect\citeauthoryear{{Line} \& {Parmentier}}{{Line} \&
  {Parmentier}}{2016}]{line_2016}
{Line} M.~R.,  {Parmentier} V.,  2016, \mn@doi [\apj]
  {10.3847/0004-637X/820/1/78}, \href
  {https://ui.adsabs.harvard.edu/abs/2016ApJ...820...78L} {820, 78}

\bibitem[\protect\citeauthoryear{{Line} et~al.,}{{Line}
  et~al.}{2016}]{2016AJ....152..203L}
{Line} M.~R.,  et~al., 2016, \mn@doi [\aj] {10.3847/0004-6256/152/6/203}, \href
  {https://ui.adsabs.harvard.edu/abs/2016AJ....152..203L} {152, 203}

\bibitem[\protect\citeauthoryear{{Lines} et~al.,}{{Lines}
  et~al.}{2018a}]{2018MNRAS.481..194L}
{Lines} S.,  et~al., 2018a, \mn@doi [\mnras] {10.1093/mnras/sty2275}, \href
  {https://ui.adsabs.harvard.edu/abs/2018MNRAS.481..194L} {481, 194}

\bibitem[\protect\citeauthoryear{{Lines} et~al.,}{{Lines}
  et~al.}{2018b}]{2018A&A...615A..97L}
{Lines} S.,  et~al., 2018b, \mn@doi [\aap] {10.1051/0004-6361/201732278}, \href
  {https://ui.adsabs.harvard.edu/abs/2018A&A...615A..97L} {615, A97}

\bibitem[\protect\citeauthoryear{{Lines}, {Mayne}, {Manners}, {Boutle},
  {Drummond}, {Mikal-Evans}, {Kohary}  \& {Sing}}{{Lines}
  et~al.}{2019}]{lines_2019}
{Lines} S.,  {Mayne} N.~J.,  {Manners} J.,  {Boutle} I.~A.,  {Drummond} B.,
  {Mikal-Evans} T.,  {Kohary} K.,   {Sing} D.~K.,  2019, \mn@doi [\mnras]
  {10.1093/mnras/stz1788}, \href
  {https://ui.adsabs.harvard.edu/abs/2019MNRAS.488.1332L} {488, 1332}

\bibitem[\protect\citeauthoryear{{Mayne} et~al.,}{{Mayne}
  et~al.}{2014}]{2014A&A...561A...1M}
{Mayne} N.~J.,  et~al., 2014, \mn@doi [\aap] {10.1051/0004-6361/201322174},
  \href {https://ui.adsabs.harvard.edu/abs/2014A&A...561A...1M} {561, A1}

\bibitem[\protect\citeauthoryear{{Mayne} et~al.,}{{Mayne}
  et~al.}{2017a}]{mayne_2017}
{Mayne} N.~J.,  et~al., 2017a, \mn@doi [\aap] {10.1051/0004-6361/201730465},
  \href {https://ui.adsabs.harvard.edu/abs/2017A&A...604A..79M} {604, A79}

\bibitem[\protect\citeauthoryear{{Mayne} et~al.,}{{Mayne}
  et~al.}{2017b}]{2017A&A...604A..79M}
{Mayne} N.~J.,  et~al., 2017b, \mn@doi [\aap] {10.1051/0004-6361/201730465},
  \href {https://ui.adsabs.harvard.edu/abs/2017A&A...604A..79M} {604, A79}

\bibitem[\protect\citeauthoryear{{Mayne}, {Drummond}, {Debras}, {Jaupart},
  {Manners}, {Boutle}, {Baraffe}  \& {Kohary}}{{Mayne}
  et~al.}{2019}]{mayne_2019}
{Mayne} N.~J.,  {Drummond} B.,  {Debras} F.,  {Jaupart} E.,  {Manners} J.,
  {Boutle} I.~A.,  {Baraffe} I.,   {Kohary} K.,  2019, \mn@doi [\apj]
  {10.3847/1538-4357/aaf6e9}, \href
  {https://ui.adsabs.harvard.edu/abs/2019ApJ...871...56M} {871, 56}

\bibitem[\protect\citeauthoryear{{Mendon{\c{c}}a}, {Malik}, {Demory}  \&
  {Heng}}{{Mendon{\c{c}}a} et~al.}{2018a}]{mendonca_2018b}
{Mendon{\c{c}}a} J.~M.,  {Malik} M.,  {Demory} B.-O.,   {Heng} K.,  2018a,
  \mn@doi [\aj] {10.3847/1538-3881/aaaebc}, \href
  {https://ui.adsabs.harvard.edu/abs/2018AJ....155..150M} {155, 150}

\bibitem[\protect\citeauthoryear{{Mendon{\c{c}}a}, {Tsai}, {Malik}, {Grimm}  \&
  {Heng}}{{Mendon{\c{c}}a} et~al.}{2018b}]{mendonca_2018}
{Mendon{\c{c}}a} J.~M.,  {Tsai} S.-m.,  {Malik} M.,  {Grimm} S.~L.,   {Heng}
  K.,  2018b, \mn@doi [\apj] {10.3847/1538-4357/aaed23}, \href
  {https://ui.adsabs.harvard.edu/abs/2018ApJ...869..107M} {869, 107}

\bibitem[\protect\citeauthoryear{{Molli{\`e}re}, {van Boekel}, {Bouwman},
  {Henning}, {Lagage}  \& {Min}}{{Molli{\`e}re}
  et~al.}{2017}]{2017A&A...600A..10M}
{Molli{\`e}re} P.,  {van Boekel} R.,  {Bouwman} J.,  {Henning} T.,  {Lagage}
  P.~O.,   {Min} M.,  2017, \mn@doi [\aap] {10.1051/0004-6361/201629800}, \href
  {https://ui.adsabs.harvard.edu/abs/2017A&A...600A..10M} {600, A10}

\bibitem[\protect\citeauthoryear{{Morley}, {Fortney}, {Marley}, {Visscher},
  {Saumon}  \& {Leggett}}{{Morley} et~al.}{2012}]{morley_2012}
{Morley} C.~V.,  {Fortney} J.~J.,  {Marley} M.~S.,  {Visscher} C.,  {Saumon}
  D.,   {Leggett} S.~K.,  2012, \mn@doi [\apj] {10.1088/0004-637X/756/2/172},
  \href {https://ui.adsabs.harvard.edu/abs/2012ApJ...756..172M} {756, 172}

\bibitem[\protect\citeauthoryear{{Morley}, {Fortney}, {Marley}, {Zahnle},
  {Line}, {Kempton}, {Lewis}  \& {Cahoy}}{{Morley} et~al.}{2015}]{morley_2015}
{Morley} C.~V.,  {Fortney} J.~J.,  {Marley} M.~S.,  {Zahnle} K.,  {Line} M.,
  {Kempton} E.,  {Lewis} N.,   {Cahoy} K.,  2015, \mn@doi [\apj]
  {10.1088/0004-637X/815/2/110}, \href
  {https://ui.adsabs.harvard.edu/abs/2015ApJ...815..110M} {815, 110}

\bibitem[\protect\citeauthoryear{{Nikolov} et~al.,}{{Nikolov}
  et~al.}{2015}]{2015MNRAS.447..463N}
{Nikolov} N.,  et~al., 2015, \mn@doi [\mnras] {10.1093/mnras/stu2433}, \href
  {https://ui.adsabs.harvard.edu/abs/2015MNRAS.447..463N} {447, 463}

\bibitem[\protect\citeauthoryear{{Parmentier}, {Showman}  \&
  {Lian}}{{Parmentier} et~al.}{2013}]{2013A&A...558A..91P}
{Parmentier} V.,  {Showman} A.~P.,   {Lian} Y.,  2013, \mn@doi [\aap]
  {10.1051/0004-6361/201321132}, \href
  {https://ui.adsabs.harvard.edu/abs/2013A&A...558A..91P} {558, A91}

\bibitem[\protect\citeauthoryear{{Parmentier}, {Fortney}, {Showman}, {Morley}
  \& {Marley}}{{Parmentier} et~al.}{2016}]{parmentier_2016}
{Parmentier} V.,  {Fortney} J.~J.,  {Showman} A.~P.,  {Morley} C.,   {Marley}
  M.~S.,  2016, \mn@doi [\apj] {10.3847/0004-637X/828/1/22}, \href
  {https://ui.adsabs.harvard.edu/abs/2016ApJ...828...22P} {828, 22}

\bibitem[\protect\citeauthoryear{{Parmentier}, {Showman}  \&
  {Fortney}}{{Parmentier} et~al.}{2020}]{parmentier_2020}
{Parmentier} V.,  {Showman} A.~P.,   {Fortney} J.~J.,  2020, arXiv e-prints,
  \href {https://ui.adsabs.harvard.edu/abs/2020arXiv201006934P} {p.
  arXiv:2010.06934}

\bibitem[\protect\citeauthoryear{{Pinhas}, {Madhusudhan}, {Gandhi}  \&
  {MacDonald}}{{Pinhas} et~al.}{2019}]{Pinhas_2019}
{Pinhas} A.,  {Madhusudhan} N.,  {Gandhi} S.,   {MacDonald} R.,  2019, \mn@doi
  [\mnras] {10.1093/mnras/sty2544}, \href
  {https://ui.adsabs.harvard.edu/abs/2019MNRAS.482.1485P} {482, 1485}

\bibitem[\protect\citeauthoryear{{Pluriel}, {Zingales}, {Leconte}  \&
  {Parmentier}}{{Pluriel} et~al.}{2020}]{Pluriel_2020}
{Pluriel} W.,  {Zingales} T.,  {Leconte} J.,   {Parmentier} V.,  2020, \mn@doi
  [\aap] {10.1051/0004-6361/202037678}, \href
  {https://ui.adsabs.harvard.edu/abs/2020A&A...636A..66P} {636, A66}

\bibitem[\protect\citeauthoryear{{Powell}, {Zhang}, {Gao}  \&
  {Parmentier}}{{Powell} et~al.}{2018}]{powell_2018}
{Powell} D.,  {Zhang} X.,  {Gao} P.,   {Parmentier} V.,  2018, \mn@doi [\apj]
  {10.3847/1538-4357/aac215}, \href
  {https://ui.adsabs.harvard.edu/abs/2018ApJ...860...18P} {860, 18}

\bibitem[\protect\citeauthoryear{{Powell}, {Louden}, {Kreidberg}, {Zhang},
  {Gao}  \& {Parmentier}}{{Powell} et~al.}{2019}]{Powell_2019}
{Powell} D.,  {Louden} T.,  {Kreidberg} L.,  {Zhang} X.,  {Gao} P.,
  {Parmentier} V.,  2019, \mn@doi [\apj] {10.3847/1538-4357/ab55d9}, \href
  {https://ui.adsabs.harvard.edu/abs/2019ApJ...887..170P} {887, 170}

\bibitem[\protect\citeauthoryear{{Rauscher} \& {Menou}}{{Rauscher} \&
  {Menou}}{2013}]{rauscher_2013}
{Rauscher} E.,  {Menou} K.,  2013, \mn@doi [\apj]
  {10.1088/0004-637X/764/1/103}, \href
  {https://ui.adsabs.harvard.edu/abs/2013ApJ...764..103R} {764, 103}

\bibitem[\protect\citeauthoryear{{Rogers} \& {Komacek}}{{Rogers} \&
  {Komacek}}{2014}]{rogers_2014}
{Rogers} T.~M.,  {Komacek} T.~D.,  2014, \mn@doi [\apj]
  {10.1088/0004-637X/794/2/132}, \href
  {https://ui.adsabs.harvard.edu/abs/2014ApJ...794..132R} {794, 132}

\bibitem[\protect\citeauthoryear{{Roman} \& {Rauscher}}{{Roman} \&
  {Rauscher}}{2019}]{roman_2019}
{Roman} M.,  {Rauscher} E.,  2019, \mn@doi [\apj] {10.3847/1538-4357/aafdb5},
  \href {https://ui.adsabs.harvard.edu/abs/2019ApJ...872....1R} {872, 1}

\bibitem[\protect\citeauthoryear{{Roman}, {Kempton}, {Rauscher}, {Harada},
  {Bean}  \& {Stevenson}}{{Roman} et~al.}{2020}]{roman_2020}
{Roman} M.~T.,  {Kempton} E. M.~R.,  {Rauscher} E.,  {Harada} C.~K.,  {Bean}
  J.~L.,   {Stevenson} K.~B.,  2020, arXiv e-prints, \href
  {https://ui.adsabs.harvard.edu/abs/2020arXiv201006936R} {p. arXiv:2010.06936}

\bibitem[\protect\citeauthoryear{{Sainsbury-Martinez}
  et~al.,}{{Sainsbury-Martinez} et~al.}{2019}]{sainsbury_martinez_2019}
{Sainsbury-Martinez} F.,  et~al., 2019, \mn@doi [\aap]
  {10.1051/0004-6361/201936445}, \href
  {https://ui.adsabs.harvard.edu/abs/2019A&A...632A.114S} {632, A114}

\bibitem[\protect\citeauthoryear{{Saumon} \& {Marley}}{{Saumon} \&
  {Marley}}{2008}]{saumon_2008}
{Saumon} D.,  {Marley} M.~S.,  2008, \mn@doi [\apj] {10.1086/592734}, \href
  {https://ui.adsabs.harvard.edu/abs/2008ApJ...689.1327S} {689, 1327}

\bibitem[\protect\citeauthoryear{{Seager} \& {Sasselov}}{{Seager} \&
  {Sasselov}}{2000}]{seager_2000}
{Seager} S.,  {Sasselov} D.~D.,  2000, \mn@doi [\apj] {10.1086/309088}, \href
  {https://ui.adsabs.harvard.edu/abs/2000ApJ...537..916S} {537, 916}

\bibitem[\protect\citeauthoryear{{Showman}, {Fortney}, {Lian}, {Marley},
  {Freedman}, {Knutson}  \& {Charbonneau}}{{Showman}
  et~al.}{2009}]{showman_2009}
{Showman} A.~P.,  {Fortney} J.~J.,  {Lian} Y.,  {Marley} M.~S.,  {Freedman}
  R.~S.,  {Knutson} H.~A.,   {Charbonneau} D.,  2009, \mn@doi [\apj]
  {10.1088/0004-637X/699/1/564}, \href
  {https://ui.adsabs.harvard.edu/abs/2009ApJ...699..564S} {699, 564}

\bibitem[\protect\citeauthoryear{{Sing}, {Vidal-Madjar}, {Lecavelier des
  Etangs}, {D{\'e}sert}, {Ballester}  \& {Ehrenreich}}{{Sing}
  et~al.}{2008}]{2008ApJ...686..667S}
{Sing} D.~K.,  {Vidal-Madjar} A.,  {Lecavelier des Etangs} A.,  {D{\'e}sert}
  J.~M.,  {Ballester} G.,   {Ehrenreich} D.,  2008, \mn@doi [\apj]
  {10.1086/590076}, \href
  {https://ui.adsabs.harvard.edu/abs/2008ApJ...686..667S} {686, 667}

\bibitem[\protect\citeauthoryear{{Sing} et~al.,}{{Sing}
  et~al.}{2016}]{2016Natur.529...59S}
{Sing} D.~K.,  et~al., 2016, \mn@doi [\nat] {10.1038/nature16068}, \href
  {https://ui.adsabs.harvard.edu/abs/2016Natur.529...59S} {529, 59}

\bibitem[\protect\citeauthoryear{{Stevenson}}{{Stevenson}}{2016}]{stevenson_2016}
{Stevenson} K.~B.,  2016, \mn@doi [\apjl] {10.3847/2041-8205/817/2/L16}, \href
  {https://ui.adsabs.harvard.edu/abs/2016ApJ...817L..16S} {817, L16}

\bibitem[\protect\citeauthoryear{{Stevenson} et~al.,}{{Stevenson}
  et~al.}{2010}]{stevenson_2010}
{Stevenson} K.~B.,  et~al., 2010, \mn@doi [\nat] {10.1038/nature09013}, \href
  {http://ads.ari.uni-heidelberg.de/abs/2010Natur.464.1161S} {464, 1161}

\bibitem[\protect\citeauthoryear{{Taylor}, {Parmentier}, {Irwin}, {Aigrain},
  {Lee}  \& {Krissansen-Totton}}{{Taylor} et~al.}{2020}]{Taylor_2020}
{Taylor} J.,  {Parmentier} V.,  {Irwin} P. G.~J.,  {Aigrain} S.,  {Lee} G.
  K.~H.,   {Krissansen-Totton} J.,  2020, \mn@doi [\mnras]
  {10.1093/mnras/staa552}, \href
  {https://ui.adsabs.harvard.edu/abs/2020MNRAS.493.4342T} {493, 4342}

\bibitem[\protect\citeauthoryear{{Tennyson} \& {Yurchenko}}{{Tennyson} \&
  {Yurchenko}}{2012}]{2012MNRAS.425...21T}
{Tennyson} J.,  {Yurchenko} S.~N.,  2012, \mn@doi [\mnras]
  {10.1111/j.1365-2966.2012.21440.x}, \href
  {https://ui.adsabs.harvard.edu/abs/2012MNRAS.425...21T} {425, 21}

\bibitem[\protect\citeauthoryear{{Tennyson} et~al.,}{{Tennyson}
  et~al.}{2016}]{2016JMoSp.327...73T}
{Tennyson} J.,  et~al., 2016, \mn@doi [Journal of Molecular Spectroscopy]
  {10.1016/j.jms.2016.05.002}, \href
  {https://ui.adsabs.harvard.edu/abs/2016JMoSp.327...73T} {327, 73}

\bibitem[\protect\citeauthoryear{{Toon}, {Turco}, {Hamill}, {Kiang}  \&
  {Whitten}}{{Toon} et~al.}{1979}]{toon_1979}
{Toon} O.~B.,  {Turco} R.~P.,  {Hamill} P.,  {Kiang} C.~S.,   {Whitten} R.~C.,
  1979, \mn@doi [Journal of Atmospheric Sciences]
  {10.1175/1520-0469(1979)036<0718:AODMDA>2.0.CO;2}, \href
  {https://ui.adsabs.harvard.edu/abs/1979JAtS...36..718T} {36, 718}

\bibitem[\protect\citeauthoryear{{Tremblin}, {Amundsen}, {Mourier}, {Baraffe},
  {Chabrier}, {Drummond}, {Homeier}  \& {Venot}}{{Tremblin}
  et~al.}{2015}]{2015ApJ...804L..17T}
{Tremblin} P.,  {Amundsen} D.~S.,  {Mourier} P.,  {Baraffe} I.,  {Chabrier} G.,
   {Drummond} B.,  {Homeier} D.,   {Venot} O.,  2015, \mn@doi [\apjl]
  {10.1088/2041-8205/804/1/L17}, \href
  {https://ui.adsabs.harvard.edu/abs/2015ApJ...804L..17T} {804, L17}

\bibitem[\protect\citeauthoryear{{Tremblin} et~al.,}{{Tremblin}
  et~al.}{2017}]{tremblin_2017}
{Tremblin} P.,  et~al., 2017, \mn@doi [\apj] {10.3847/1538-4357/aa6e57}, \href
  {https://ui.adsabs.harvard.edu/abs/2017ApJ...841...30T} {841, 30}

\bibitem[\protect\citeauthoryear{{Turco}, {Hamill}, {Toon}, {Whitten}  \&
  {Kiang}}{{Turco} et~al.}{1979}]{turco_1979}
{Turco} R.~P.,  {Hamill} P.,  {Toon} O.~B.,  {Whitten} R.~C.,   {Kiang} C.~S.,
  1979, \mn@doi [Journal of Atmospheric Sciences]
  {10.1175/1520-0469(1979)036<0699:AODMDA>2.0.CO;2}, \href
  {https://ui.adsabs.harvard.edu/abs/1979JAtS...36..699T} {36, 699}

\bibitem[\protect\citeauthoryear{{Visscher}, {Lodders}  \& {Fegley}}{{Visscher}
  et~al.}{2006}]{visscher_2006}
{Visscher} C.,  {Lodders} K.,   {Fegley} Bruce J.,  2006, \mn@doi [\apj]
  {10.1086/506245}, \href
  {https://ui.adsabs.harvard.edu/abs/2006ApJ...648.1181V} {648, 1181}

\bibitem[\protect\citeauthoryear{{Visscher}, {Lodders}  \& {Fegley}}{{Visscher}
  et~al.}{2010}]{visscher_2010}
{Visscher} C.,  {Lodders} K.,   {Fegley} Bruce J.,  2010, \mn@doi [\apj]
  {10.1088/0004-637X/716/2/1060}, \href
  {https://ui.adsabs.harvard.edu/abs/2010ApJ...716.1060V} {716, 1060}

\bibitem[\protect\citeauthoryear{{Wakeford} et~al.,}{{Wakeford}
  et~al.}{2017}]{Wakeford_2017}
{Wakeford} H.~R.,  et~al., 2017, \mn@doi [Science] {10.1126/science.aah4668},
  \href {https://ui.adsabs.harvard.edu/abs/2017Sci...356..628W} {356, 628}

\bibitem[\protect\citeauthoryear{{Wood} et~al.,}{{Wood}
  et~al.}{2014}]{2014QJRMS.140.1505W}
{Wood} N.,  et~al., 2014, \mn@doi [Quarterly Journal of the Royal
  Meteorological Society] {10.1002/qj.2235}, \href
  {https://ui.adsabs.harvard.edu/abs/2014QJRMS.140.1505W} {140, 1505}

\bibitem[\protect\citeauthoryear{{Zellem} et~al.,}{{Zellem}
  et~al.}{2014}]{zellem_2014}
{Zellem} R.~T.,  et~al., 2014, \mn@doi [\apj] {10.1088/0004-637X/790/1/53},
  \href {https://ui.adsabs.harvard.edu/abs/2014ApJ...790...53Z} {790, 53}

\bibitem[\protect\citeauthoryear{{Zhang} \& {Showman}}{{Zhang} \&
  {Showman}}{2018}]{2018ApJ...866....2Z}
{Zhang} X.,  {Showman} A.~P.,  2018, \mn@doi [\apj] {10.3847/1538-4357/aada7c},
  \href {https://ui.adsabs.harvard.edu/abs/2018ApJ...866....2Z} {866, 2}

\makeatother
\end{thebibliography}

\appendix

\section{Condensate Properties}
\label{app:sat_vap}
\begin{table*}
\caption{Condensate Properties }
\label{Tbl:SVP}
\begin{tabular}{lccccc}
\hline
Species  & $q_\mathrm{below}$ [g/g] & $\mu_\text{c}$ [$\mathrm{g\, mol^{-1}}$] & $\rho_\mathrm{c}$ [$\mathrm{g\, cm^{-3}}$] & $P_\mathrm{svp}$ [bar] & $P_\mathrm{svp}$ Source \\
\hline
$\mathrm{Al_2O_3}$ & $1.1\times 10^{-4}$ & 101.961 & 3.987 & $e^{22.01 -73503\, \mathrm{K}/T}$ & \citet{kozasa_1989} \\
$\mathrm{Fe}$ & $4.48\times 10^{-4}$ & 55.845 & 7.875 & $10^{7.09-20833\,\mathrm{K}/T}$ & \citet{visscher_2010} \\
$\mathrm{Na_2S}$ & $5.32\times 10^{-5}$ & 78.05 &  1.856 & $10^{8.5497-13889\, \mathrm{K}/T}$ &  \citet{visscher_2006} \\
$\mathrm{NH_3}$ & $4.48\times 10^{-4}$ & 17.0 & 0.84&  $e^{-86596\, \mathrm{K}^2/T^2 - 2161 \mathrm{K}/T + 10.53}$ & \citet{2001ApJ...556..872A}  \\
$\mathrm{KCl}$ & $6.1\times 10^{-6}$ & 74.5 & 1.99 & $10^{7.6106 - 11382\,\mathrm{K}/T}$ & \citet{morley_2012} \\
$\mathrm{MnS}$ & $2.53\times 10^{-5}$ & 87.00 & 4.0 & $10^{11.5315-23810\,\mathrm{K}/T}$ & \citet{visscher_2006}  \\
$\mathrm{ZnS}$ & $3.72\times 10^{-6}$ & 97.46 & 4.04 & $10^{12.8117-15873\,\mathrm{K}/T}$ & \citet{visscher_2006} \\
$\mathrm{Cr}$ & $1.77\times 10^{-5}$ & 51.996 & 7.15 & $10^{7.2688-20353\,\mathrm{K}/T}$ & See text \\
$\mathrm{MgSiO_3}$ & $1.55\times 10^{-3}$ & 100.4 &  3.192 & $10^{11.83 - 27250\,\mathrm{K}/T}$ &  \cite{visscher_2010} \\
$\mathrm{Mg_2SiO_4}$ & $1.09\times 10^{-3}$ & 140.7 & 3.214 & $10^{-32488\,\mathrm{K}/T + 14.88 - 0.2\log_{10}P}$ & \cite{visscher_2010} \\
\hline
\end{tabular}
\end{table*}

As there does not exist a central repository for the {\sc EddySed} code and multiple versions exist, we catalogue here the saturation vapour pressures and condensate properties used in our version of the code (see Table \ref{Tbl:SVP}).   A discussion of most the saturation vapour pressures and their incorporation in the {\sc EddySed} code can be found in \citet{2001ApJ...556..872A} and \citet{morley_2012}.  An exception is the saturation vapour pressure for $\mathrm{Cr}$ which, according to comments in the {\sc EddySed} source code, was obtained from a private communication.  It is similar to the saturation vapour pressure for $\mathrm{Cr}$ used in \citet{morley_2012}; however, they differ by up to 21\% for temperatures between $1000\,\mathrm{K}$ and $2000\,\mathrm{K}$.  We also note that in cases where the prescribed saturation vapour pressure has a dependence on the metallicity, the metallicity is taken to be $0.0$.

% Don't change these lines
\bsp	% typesetting comment
\label{lastpage}
\end{document}